\documentclass[11pt,a4paper]{article}

\usepackage{jheppub}
\usepackage{amsmath}
\usepackage{amssymb}
\usepackage{graphicx}
\usepackage{dcolumn}
\usepackage{bm}
\usepackage{tabularx}
\usepackage[titles]{tocloft}
\usepackage{color}
\usepackage{url}
\usepackage{xspace}
\usepackage{csquotes}
\usepackage{xcolor}

\newcommand{\append}[1]{\protect\refstepcounter{section}
                \section*{Appendix \thesection \, #1}
                \addcontentsline{toc}{section}{Appendix \thesection: #1}}

\def\MIN{{\sc Min}}
\def\MED{{\sc Med}}
\def\MAX{{\sc Max}}

\newcommand{\pbar}{\bar{p}}

\newcommand{\wimp}{\textsc{WIMP}}
\newcommand{\wimps}{\textsc{WIMPs}}

\mathchardef\mhyphen="2D

\newlength{\dhatheight}

\newcommand{\sigmav}{\ensuremath{\langle \sigma v \rangle}\xspace}

\providecommand\physrep{\ref@jnl{Phys.~Rep.}}%
\providecommand\apjs{\ref@jnl{ApJS}}%
\providecommand{\jcap}{\ref@jnl{JCAP}}%

\begin{document}

\begin{flushright}
CERN-TH-2018-149
\end{flushright}

\begin{center}
\Large\bf\boldmath
\vspace*{2.0cm}SuperIso Relic v4: A program for calculating dark matter \\
and flavour physics observables in Supersymmetry
\unboldmath
\end{center}

\vspace{0.8cm}
\begin{center}
A. Arbey\footnote{Electronic address: \tt alexandre.arbey@ens-lyon.fr}$^,$\footnote{Also Institut Universitaire de France, 103 boulevard Saint-Michel, 75005 Paris, France}$^{,a,b}$, F. Mahmoudi$^{\dagger,}$\footnote{Electronic address: \tt mahmoudi@in2p3.fr}$^{,a,b}$, G. Robbins\footnote{Electronic address: \tt glenn.robbins@univ-lyon1.fr}$^{,c}$\\[0.4cm]
{$^a$\sl Univ Lyon, Univ Lyon 1, CNRS/IN2P3, Institut de Physique Nucl\'eaire de Lyon UMR5822, F-69622 Villeurbanne, France.\\[0.2cm]}
{$^b$\sl CERN, Theoretical Physics Department, CH-1211 Geneva 23, Switzerland.\\[0.2cm]}
{$^c$\sl Univ Lyon, Univ Lyon 1, ENS de Lyon, CNRS, Centre de Recherche Astrophysique de Lyon UMR5574, F-69230 Saint-Genis-Laval, France.}

\end{center}
\vspace{0.9cm}
\begin{abstract}
\noindent We describe \texttt{SuperIso Relic}, a public program for the calculation of dark matter relic density and direct and indirect detection rates, which includes in addition the {\tt SuperIso} routines for the calculation of flavour physics observables. \texttt{SuperIso Relic v4} incorporates many new features, namely the possibility of multiprocessor calculation of the relic density, new cosmological models, and the implementations of the calculation of the observables related to direct and indirect detection experiments.
Furthermore, the new version includes an implementation of the nuclear and astrophysical uncertainties, from namely nuclear form factors, dark matter density and velocity, as well as cosmic ray propagation through the galactic medium.\\
\end{abstract}
\newpage
\tableofcontents
%
\newpage
\section{Introduction} 

{\tt SuperIso Relic} is a mixed C / Fortran program devoted to the calculation of dark matter and flavour observables in supersymmetry (SUSY). It is an extension of {\tt SuperIso} \cite{Mahmoudi:2007vz,Mahmoudi:2008tp,Mahmoudi:2009zz} which contains the calculation of the flavour observables, and includes additional routines from {\tt AlterBBN} \cite{Arbey:2011nf,Arbey:2018zfh}. The previous versions of {\tt SuperIso Relic} were devoted to the calculation of the relic density in standard and modified cosmological scenarios \cite{Arbey:2009gu,Arbey:2011zz}. 

Many new features have been added in the current version of {\tt SuperIso Relic}, which justifies a major version number change. The main features are the possibility of multiprocessor calculation of the relic density using the {\tt OpenMP} library, new cosmological models, and the implementations of the calculation of observables related to direct and indirect detection experiments, with a focus on the possibility to alter the astrophysical description of dark matter and propagation of dark matter annihilation products. 
These features distinguish {\tt SuperIso Relic} from the other existing public codes such as {\tt DarkSusy} \cite{Gondolo:2004sc}, {\tt micrOMEGAs} \cite{Belanger:2004yn}, {\tt MadDM} \cite{Ambrogi:2018jqj} and {\tt DarkBit} \cite{Workgroup:2017lvb,Athron:2017ard}.

{\tt SuperIso Relic} is public and open-source, and is released with the GPL version 3 licence. A detailed description of the code is provided in the following; the content of the SuperIso Relic package is described in section~\ref{content}, the compilation and installation instructions are provided in section~\ref{compilation}, the input and output descriptions are given in section~\ref{sample}, and some illustrative results are presented in section~\ref{result}. The appendices contain the description of the cosmological scenarios as well as the calculation of indirect and direct detection rates.

\section{Content of the SuperIso Relic package}%
\label{content}
\subsection{General description}

The main directory of {\tt SuperIso Relic} contains 23 main programs in C language. {\tt sm.c}, {\tt sm\_chi2.c}, {\tt flha.c}, , {\tt slha\_chi2.c} and {\tt thdm.c} are only relevant for {\tt SuperIso} and do not involve dark matter observables. The other programs relies on the SUSY Les Houches Accord (SLHA) format \cite{Skands:2003cj,Allanach:2008qq} as input format for supersymmetric models. 10 programs are common to both {\tt SuperIso} and {\tt SuperIso Relic}:
\begin{itemize}
 \item {\tt amsb.c}: calculation of observables for an AMSB model point.
 \item {\tt cmssm.c}: calculation of the observables for a CMSSM model point.
 \item {\tt cnmssm.c}: calculation of the observables for a semi constrained NMSSM model point.
 \item {\tt gmsb.c}: calculation of the observables for a GMSB model point.
 \item {\tt hcamsb.c}: calculation of the observables for a hypercharged AMSB model point.
 \item {\tt mmamsb.c}: calculation of the observables for a mixed moduli AMSB model point.
 \item {\tt nnuhm.c}: calculation of the observables for a NMSSM NUHM model point.
 \item {\tt ngmsb.c}: calculation of the observables for a NMSSM GMSB model point.
 \item {\tt nuhm.c}: calculation of the observables for a NUHM model point.
 \item {\tt slha.c}: scan of the user provided SLHA file and calculation of the observables.
\end{itemize}
and 8 programs are {\tt SuperIso Relic}-specific:
\begin{itemize}
\item {\tt create\_propagation.c}: calculation of AMS-02 antiproton constraints \cite{Aguilar:2016kjl} in a user-specified propagation model and profile.
\item {\tt direct.c}: calculation of the direct detection observables, incorporating the limits by PANDAX-2 \cite{Cui:2017nnn}, XENON1T \cite{Aprile:2017iyp} and PICO60 \cite{Amole:2017dex}.
\item {\tt indirect.c}: calculation of the indirect detection observables, incorporating the photon results of FERMI-LAT \cite{Ackermann:2015zua} and AMS-02 antiproton limits \cite{Aguilar:2016kjl}.
\item {\tt test\_phi.c}: scans the user provided SLHA file and calculates the relic density in a scenario with a decaying cosmological scalar field.
\item {\tt test\_modeleff.c}: scans the user provided SLHA file and calculates the relic density for different radiation equations of state.
\item {\tt test\_standmod.c}: scans the user provided SLHA file and calculates the relic density in a model with dark density and dark entropy in the early Universe. The BBN constraints are checked automatically with {\tt AlterBBN}.
\item {\tt test\_reheating.c}: scans the user provided SLHA file and calculates the relic density in a model with dark density and reheating in the early Universe. The BBN constraints are checked automatically with {\tt AlterBBN}.
\item {\tt test\_widthcalc.c}: scans the user provided SLHA file and calculates the relic density using the different width calculators which are included.
\end{itemize}
The files \texttt{README} and \texttt{README.superiso} describe {\tt SuperIso} and {\tt SuperIso Relic}, respectively. \texttt{example.lha} is an example of SLHA file.
The files \texttt{configure} and \texttt{Makefile} allow the user to configure and compile the package.
The \texttt{src/} directory contains the source files of the package, and in particular the C files:
\begin{itemize}
 \item \texttt{bbn.c}: calculation of BBN constraints using \texttt{AlterBBN}.
 \item \texttt{bbnrate.c}: nuclear rates for \texttt{AlterBBN}.
 \item \texttt{DDetection.c}: calculation of dark matter direct detection observables.
 \item \texttt{DDhalo.c}: routines related to the dark matter halo profile for direct detection.
 \item \texttt{DDMSSM.c}: routines related to the MSSM for direct detection.
 \item \texttt{DDnuclear.c}: routines related to the nuclear models for direct detection.
 \item \texttt{feynhiggs.c}: procedures calling \texttt{FeynHiggs}.
 \item \texttt{hdecay.c}: procedure calling \texttt{Hdecay}.
 \item \texttt{IDetection.c}: calculation of indirect detection observables.
 \item \texttt{omega.c}: calculation of the dark matter relic density.
 \item \texttt{propagation.c} and \texttt{propagation.h}: routines for the proton/antiproton propagation model.
 \item \texttt{sdecay.c}: procedure calling \texttt{Sdecay}/\texttt{SUSYHIT}.
 \item \texttt{widthcalc.c}: routines related to the Higgs width calculation.
 \item \texttt{include.h}: definitions and prototypes.
 \item \texttt{include\_dm.h}: definitions and prototypes specific to \texttt{SuperIso Relic}.
\end{itemize}
and five directories:
\begin{itemize}
 \item \texttt{contrib/}: contains external programs useful for the relic density (\texttt{FeynHiggs}, \texttt{Hdecay}, \texttt{SUSYHIT}).
 \item \texttt{sgStar\_heff/}: contains different tables of $g_{\rm eff}(T)$ and $h_{\rm eff}(T)$.
 \item \texttt{relic/}: contains all the Fortran routines used for the calculation of the relic density.
 \item \texttt{gammas/}: contains the data for interpreting the FERMI-LAT results.
 \item \texttt{antiprotons/}: contains the data for interpreting the AMS-02 results.
\end{itemize}

In addition, routines interfacing {\tt SuperIso Relic} with \texttt{ISAJET} \cite{Paige:2003mg}, \texttt{SOFTSUSY} \cite{Allanach:2001kg}, \texttt{SPheno} \cite{Porod:2003um}, \texttt{SuSpect} \cite{Djouadi:2002ze} and \texttt{NMSSMTools} \cite{Ellwanger:2009dp} are provided. \texttt{FeynHiggs} \cite{Heinemeyer:1998yj} or \texttt{Hdecay}, \texttt{Sdecay} and \texttt{SUSYHIT} \cite{Djouadi:1997yw,Muhlleitner:2003vg,Djouadi:2006bz} are used to compute the decay widths of the Higgs bosons and supersymmetric particles useful for the relic density and neutralino annihilation cross section calculations. The propagation model included in \texttt{propagation.c} is the one used in \cite{Boudaud:2014qra, Giesen:2015ufa}.

The directory \texttt{relic/} contains the routines necessary for the relic density calculation. \texttt{Weff/} and \texttt{squared/} implement 3011 annihilation and coannihilation processes for the MSSM and \texttt{WeffN/} and \texttt{squaredn/} 4219 processes for the NMSSM in Fortran code. In comparison to the previous versions of {\tt SuperIso Relic}, all these files have been re-generated with \texttt{FeynArts}/\texttt{FormCalc} \cite{Hahn:2000jm,Hahn:2006qw,Hahn:2016ebn} using the \texttt{LANHEP} \cite{Semenov:1998eb,Semenov:2008jy} model files, in order to improve the compilation speed as well as to enable the use of the \texttt{OpenMP} library to compute the relic density. All the \texttt{COMMON} blocks which are automatically generated by \texttt{FormCalc} are defined as \texttt{THREADPRIVATE} with respect to \texttt{OpenMP}. The relic density Fortran libraries can be compiled as static or as dynamic, and the interface is insured by \texttt{staticlibcalc.c}/\texttt{staticlibcalcN.c} and \texttt{dynlibcalc.c}, respectively. \texttt{model\_superiso.F} enables the exchange of the supersymmetric parameters between the C and Fortran parts of the code.

\subsection{Parameter structures}

\texttt{SuperIso Relic} uses several C-structures to exchange data between the different subroutines, which are described in the following.

\subsubsection{Internal parameters}

The \texttt{parameters} structure contains the central parameters of \texttt{SuperIso Relic}, which are exchanged between most of the routines.

\begin{verbatim}
typedef struct parameters
/* structure containing all the scanned parameters from the SLHA file */
{
int SM;
int model; /* CMSSM=1, GMSB=2, AMSB=3 */
int generator; /* ISAJET=1, SOFTSUSY=3, SPHENO=4, SUSPECT=5, NMSSMTOOLS=6 */
double Q; /* Qmax ; default = M_EWSB = sqrt(m_stop1*mstop2) */

double m0,m12,tan_beta,sign_mu,A0; /* CMSSM parameters */
double Lambda,Mmess,N5,cgrav,m32; /* AMSB, GMSB parameters */
double mass_Z,mass_W,mass_b,mass_top_pole,mass_tau_pole; /* SM parameters */
double inv_alpha_em,alphas_MZ,Gfermi,GAUGE_Q; /* SM parameters */
double charg_Umix[3][3],charg_Vmix[3][3],stop_mix[3][3],sbot_mix[3][3],stau_mix[3][3]
,neut_mix[6][6],mass_neut[6],alpha; /* mass mixing matrices */
double Min,M1_Min,M2_Min,M3_Min,At_Min,Ab_Min,Atau_Min,M2H1_Min,M2H2_Min,mu_Min,
M2A_Min,tb_Min,mA_Min; /* optional input parameters at scale Min */
double MeL_Min,MmuL_Min,MtauL_Min,MeR_Min,MmuR_Min,MtauR_Min; /* optional input 
parameters at scale Min */
double MqL1_Min,MqL2_Min,MqL3_Min,MuR_Min,McR_Min,MtR_Min,MdR_Min,MsR_Min,
MbR_Min; /* optional input parameters at scale Min */
double N51,N52,N53,M2H1_Q,M2H2_Q; /* optional input parameters (N51...3: GMSB) */
double mass_d,mass_u,mass_s,mass_c,mass_t,mass_e,mass_nue,mass_mu,mass_num,mass_tau,
mass_nut; /* SM masses */
double mass_gluon,mass_photon,mass_Z0; /* SM masses */
double mass_h0,mass_H0,mass_A0,mass_H,mass_dnl,mass_upl,mass_stl,mass_chl,mass_b1,
mass_t1; /* Higgs & superparticle masses */
double mass_el,mass_nuel,mass_mul,mass_numl,mass_tau1,mass_nutl,mass_gluino,mass_cha1,
mass_cha2; /* superparticle masses */
double mass_dnr,mass_upr,mass_str,mass_chr,mass_b2,mass_t2,mass_er,mass_mur,
mass_tau2; /* superparticle masses */
double mass_nuer,mass_numr,mass_nutr,mass_graviton,
mass_gravitino; /* superparticle masses */
double gp,g2,gp_Q,g2_Q,g3_Q,YU_Q,yut[4],YD_Q,yub[4],YE_Q,yutau[4]; /* couplings */
double HMIX_Q,mu_Q,tanb_GUT,Higgs_VEV,mA2_Q,MSOFT_Q,M1_Q,M2_Q,M3_Q; /* parameters
at scale Q */
double MeL_Q,MmuL_Q,MtauL_Q,MeR_Q,MmuR_Q,MtauR_Q,MqL1_Q,MqL2_Q,MqL3_Q,MuR_Q,McR_Q,MtR_Q,
MdR_Q,MsR_Q,MbR_Q; /* masses at scale Q */
double AU_Q,A_u,A_c,A_t,AD_Q,A_d,A_s,A_b,AE_Q,A_e,A_mu,A_tau; /* trilinear couplings */

/* SLHA2 */
int NMSSM,RV,CPV,FV;
double CKM_lambda,CKM_A,CKM_rhobar,CKM_etabar;
double PMNS_theta12,PMNS_theta23,PMNS_theta13,PMNS_delta13,PMNS_alpha1,PMNS_alpha2;
double lambdaNMSSM_Min,kappaNMSSM_Min,AlambdaNMSSM_Min,AkappaNMSSM_Min,lambdaSNMSSM_Min,
xiFNMSSM_Min,xiSNMSSM_Min,mupNMSSM_Min,mSp2NMSSM_Min,mS2NMSSM_Min,mass_H03,mass_A02,
NMSSMRUN_Q,lambdaNMSSM,kappaNMSSM,AlambdaNMSSM,AkappaNMSSM,lambdaSNMSSM,xiFNMSSM,
xiSNMSSM,mupNMSSM,mSp2NMSSM,mS2NMSSM; /* NMSSM parameters */
double PMNSU_Q,CKM_Q,IMCKM_Q,MSE2_Q,MSU2_Q,MSD2_Q,MSL2_Q,MSQ2_Q,TU_Q,TD_Q,TE_Q;
double CKM[4][4],IMCKM[4][4]; /* CKM matrix */
double H0_mix[4][4],A0_mix[4][4]; /* Higgs mixing matrices */
double sU_mix[7][7],sD_mix[7][7],sE_mix[7][7], sNU_mix[4][4]; /* mixing matrices */
double sCKM_msq2[4][4],sCKM_msl2[4][4],sCKM_msd2[4][4],sCKM_msu2[4][4],
sCKM_mse2[4][4]; /* super CKM matrices */
double PMNS_U[4][4]; /* PMNS mixing matrices */
double TU[4][4],TD[4][4],TE[4][4]; /* trilinear couplings */

/* non-SLHA*/
double mass_c_pole,mass_b_1S,mass_b_pole,mtmt;
int scheme_c_mass;
double Lambda3,Lambda4,Lambda5,Lambda6; /* Lambda QCD */
double alphasMZ_Lambda3,alphasMZ_Lambda4,alphasMZ_Lambda5,
alphasMZ_Lambda6; /* Lambda QCD */
\end{verbatim}
\[
\huge \bf \vdots
\]
\begin{verbatim}
/* Decay widths */
int widthcalc; /* 0=none, 1=hdecay, 2=feynhiggs */
double width_h0,width_H0,width_A0,width_H,width_Z,width_W,width_top,width_H03,
width_A02;
double width_gluino,width_t1,width_t2,width_b1,width_b2,width_ul,width_ur,width_dl,
width_dr;
double width_cl,width_cr,width_sl,width_sr,width_el,width_er,width_ml,width_mr,
width_tau1,width_tau2,width_gravitino;
double width_nuel,width_numl,width_nutaul,width_c1,width_c2,width_o1,width_o2,
width_o3,width_o4,width_o5;
                        
/* 2HDM */
int THDM_model;
double lambda_u[4][4],lambda_d[4][4],lambda_l[4][4];

/* NMSSMTools */
int NMSSMcoll,NMSSMtheory,NMSSMups1S,NMSSMetab1S;
        
/* SDECAY */
double BRtbW,BRtbH,BRtt1o1,BRtt1o2,BRtt1o3,BRtt1o4,BRtt2o1,BRtt2o2,BRtt2o3,BRtt2o4;
double BRgluinot1tbar,BRgluinot1bart,BRgluinodldbar,BRgluinodlbard,BRgluinodrdbar,
\end{verbatim}
\[
\huge \bf \vdots
\]
\begin{verbatim}
/* HDECAY & FeynHiggs */
double mass_h0SM,width_h0SM;
double mass_H0SM,width_H0SM;
double mass_A0SM,width_A0SM;
double BRh0bb_SM,BRh0tautau_SM,BRh0WW_SM,BRh0gg_SM,BRh0gaga_SM,BRh0ZZ_SM;
double BRH0bb_SM,BRH0tautau_SM,BRH0WW_SM,BRH0gg_SM,BRH0gaga_SM,BRH0ZZ_SM;
double BRA0bb_SM,BRA0tautau_SM,BRA0WW_SM,BRA0gg_SM,BRA0gaga_SM,BRA0ZZ_SM;

double BRh0bb,BRh0tautau,BRh0WW,BRh0gg,BRh0gaga,BRh0ZZ;
\end{verbatim}
\[
\huge \bf \vdots
\]
\begin{verbatim}
}
parameters;
\end{verbatim}

\subsubsection{Relic density}

The main relic density structure has been expanded since the previous versions of \texttt{SuperIso Relic}, to incorporate new cosmological scenarios and more parameters related to \texttt{AlterBBN}.

\begin{verbatim}
typedef struct relicparam
{
    int entropy_model,energy_model;
    double dd0,ndd,Tdend,Tddeq; /* dark density */
    double sd0,nsd,Tsend; /* dark entropy */
    double Sigmad0,nSigmad,TSigmadend; /* dark entropy injection */
    double Sigmarad0,nSigmarad,TSigmaradend; /* standard entropy injection */
    double nt0,nnt,Tnend; /* non-thermal production of relics */
    
    double quintn2,quintn3,quintn4,quintT12,quintT23,quintT34; /* quintessence */

    int phi_model; /* decaying scalar field model switch */
    double eta_phi,Gamma_phi,rhot_phi_Tmax,rho_phi; /* eta_phi = b / m_phi */
    double rhot_phi0,Tphi0;
    double T_RH;
    double Sigmatildestar;
    double Sigmatildestar_max;
    double Tstdstar_max;

    double mgravitino; /* gravitino mass */
            
    double relicmass;
    int scalar;
    
    int solver; /* switch for linear or logarithmic differential equation solver */
    
    double T; /* Temperature in GeV */
    double Y; /* Y=n/s */
    double Tfo,Tmax; /* Freeze out and maximal temperature */
    
    int full_comput; /* Switch to deactivate the fast freeze out temperature
    determination */
  
    double table_eff[276][3]; /* Reads values from the SgStar files */
   
    int use_table_rhoPD;
    double table_rhoPD[2][NTABMAX];
    int size_table_rhoPD;

    /*---------------------*/
    /* AlterBBN parameters */
    /*---------------------*/
    
    int err;
    int failsafe;
    double eta0;
    double Nnu;
    double dNnu;
    double life_neutron,life_neutron_error;
    double xinu1,xinu2,xinu3;
    double m_chi;
    double g_chi;
    double Tinit;
    int wimp;
    int SMC_wimp;
    int selfConjugate;
    int fermion;
    int EM_coupled, neut_coupled, neuteq_coupled;
    double chi2;
    int nobs;    
}
relicparam;
\end{verbatim}

For the description of the specific {\tt AlterBBN} parameters we refer the reader to \cite{Arbey:2011nf,Arbey:2018zfh}.

\subsubsection{Direct detection}

\begin{verbatim}
struct DDparameters
{
/* error option =1 to consider the underlying uncertainties, 0 otherwise */
    int nucleonSIerror; /* Spin-independent nucleon form factor */
    int nucleonSDerror; /* Spin-dependent nucleon form factor */
    int nuclearSDerror; /* Spin-dependent nuclear structure factor */
    int rho0error; /* local DM density */
    int vescerror; /* escape velocity */
    int vroterror; /* disk rotational velocity */
/* detector */
    int niso; /* isotope number */
    int A[NISOMAX]; /* proton+neutron number */
    int Z[NISOMAX]; /* proton number */
    double MA[NISOMAX]; /* isotope masses */
    double mu[NISOMAX]; /*isotope-LSP reduced mass */ 
    double massfract[NISOMAX]; /* isotope mass fraction */
    double J[NISOMAX]; /* spin */
/* LSP */
    double relicmass;
    double ASI[6]; /* quark contributions to the SI scattering amplitude */
    double  ASD[6]; /* quark contributions to the SD scattering amplitude */
    double ASDp; /* proton-LSP SD scattering amplitude */
    double ASDn; /* neutron-LSP SD scattering amplitude */
    double ASIp; /* proton-LSP SI scattering amplitude */
    double ASIn; /* neutron-LSP SI scattering amplitude */
    double ddpSI; /* proton-LSP SI scattering cross section */
    double ddnSI; /* neutron-LSP SI scattering cross section */
    double ddpSD; /* proton-LSP SD scattering cross section */
    double ddnSD; /* neutron-LSP SD scattering cross section */
/* Nucleon form factors */
/* Spin independent */
    double sigmapin[2]; /* light quark content of nucleons */
    double sigmas[2]; /* strange quark content of nucleons */
    double z[2]; /* z parameter occurring in the form factor calculation */
    double fdSIp; /* d-quark SI form factor in proton */
    double fuSIp; /* u-quark SI form factor in proton */
    double fsSIp; /* s-quark SI form factor in proton */
    double fdSIn; /* d-quark SI form factor in neutron */
    double fuSIn; /* u-quark SI form factor in neutron */
    double fsSIn; /* s-quark SI form factor in neutron */
/* spin dependent */
    double a3[2]; /* fuSDp-fdSDp */
    double a8[2]; /* fuSDp+fdSDp-2*fsSDp */
    double fsSDp[2]; /* s-quark SD form factor in proton */
    double fdSDp; /* d-quark SD form factor in proton */
    double fuSDp; /* u-quark SD form factor in proton */
    double fdSDn; /* d-quark SD form factor in neutron */
    double fuSDn; /* u-quark SD form factor in neutron */
    double fsSDn; /* s-quark SD form factor in neutron */
/* Nuclei form factors */
/* Spin independent */
    double (*FFSI)(double , int ); /* SI nuclear form factor (Helm by default) */
/* Spin dependent */
    double (*S00[NISOMAX])(double); /* SD nuclear structure factor S00 */
    double (*S00min[NISOMAX])(double); /* S00 lower bound  */
    double (*S00max[NISOMAX])(double); /* S00 upper bound */
    double (*S01[NISOMAX])(double); /* SD nuclear structure factor S01 */
    double (*S01min[NISOMAX])(double); /* S01 lower bound */
    double (*S01max[NISOMAX])(double); /* S01 upper bound */
    double (*S11[NISOMAX])(double); /* SD nuclear structure factor S11 */
    double (*S11min[NISOMAX])(double); /* S11 lower bound */
    double (*S11max[NISOMAX])(double); /* S11 upper bound */
/* DM halo */
    double rho0[2]; /* local DM density */
    double vrot[2]; /* disk rotational velocity */
    double vesc[2]; /* escape velocity */
    double vearth[3]; /* peculiar velocity of Earth in the local standard of rest */
    double vearthmed; /* Earth velocity in the galactic frame */
    double vearthmin; /* Minimal Earth velocity with regards to vrot uncertainties */
    double vearthmax; /* Maximal Earth velocity with regards to vrot uncertainties */
    double Maxwellnorm; /* Norm of the Maxwell distribution */
    double Maxwellnormmin; /* Minimal value of Maxwellnorm 
                            with regards to vrot and vesc uncertainties */
    double Maxwellnormmax; /* Maximal value */
    double(*eta)(struct DDparameters* ,double, char*); /* mean inverse speed */
};
\end{verbatim}

\subsubsection{Indirect detection}

Several structures are used for indirect detection.

\paragraph{Structure for the dwarf spheroidal galaxy list for FERMI-LAT:}
\begin{verbatim}
struct dSph{ 
    char name[500]; /* name of the galaxy */
    double J; /* J-factor */
    double deltaJ; /* J-factor uncertainty */
    double likelihood[600][3];/* tabulated likelihood */
    double likelihood_noDM; /* likelihood in the background-only hypothesis */
    int sample; /* sample to which belongs the galaxy
                   (=-1 for conservative, =0 for nominal, =1 for inclusive) */
} ;
\end{verbatim}

\paragraph{Structure for FERMI-LAT:}
\begin{verbatim}
struct fermi
{
    int NdSphs; /* number of galaxies */
    struct dSph dSphs[50]; /* list of the galaxies */
    int Nenergybins; /* number of energy bin */
    double eflux[24]; /* photon flux from DM annihilation */
    double Emin[24]; /* lower bounds of the energy bins */
    double Emax[24]; /* upper bounds of the energy bins */
};
\end{verbatim}

\paragraph{Structure for the dark matter halo:}
\begin{verbatim}
struct Structure_Halo
{
    char name[80]; /* name of the halo */
    double (*profile)(double, double); /* DM halo profile */
    double r_earth ; /* distance of the Earth from the galactic center */
    double rho_chi_solar; /* DM local density */
};
\end{verbatim}

\paragraph{Structure for the propagation parameters:}
\begin{verbatim}
struct propagation_parameters
{
    struct Structure_Propagation Propagation; /* contains the propagation model */
    struct Structure_Halo Halo; /* contains the DM halo model */
    double chi2_noDM; /* chi^2  for the background only hypothesis */
    struct array bkg; /* secondary antiproton tabulated spectra */
}
\end{verbatim}

\paragraph{Structure for the particles:}
\begin{verbatim}
struct particle
{
    char* name; 
    double mass;
    int type; /* scalar, fermion or vector */
};
\end{verbatim}

\paragraph{Structure for the processes related to indirect detection:}
\begin{verbatim}
struct process
{ 
/* defines a 2->2 process particle1 particle2 -> particle3 particle4
   or a decay particle1 -> particle3 particle4 */
  
    char* name; /* name of the process */
    struct particle particle1;
    struct particle particle2;
    struct particle particle3;
    struct particle particle4;
    int sf34; /* =1 if part 3 and 4 are different, =2 if they are the same */
    int hel; /* defines the spin of the particles */
    double sigmav; /* thermal averaged annihilation cross section 
                                        or decay branching ratio */
};
\end{verbatim}

\paragraph{Structure for the list of processes related to indirect detection:}
\begin{verbatim}
struct processes
{
    int nproc; /* number of processes */
    struct process list[100]; /* processes list */
    int dof; /* number of processes which contributes
                to > 1% to the total annihilation cross section */
    double relicmass;
    double sigmav; /* total annihilation cross section */
};
\end{verbatim}

\paragraph{Main indirect detection structure:}
\begin{verbatim}
struct IDparameters
{
    struct processes o1o1; /* neutralino annihilation processes */
    struct processes h0; /* Light CP-even Higgs decay */
    struct processes H0; /* Heavy CP-even Higgs decay */
    struct processes A0; /* CP-odd Higgs decay */
    struct processes h3; // Third CP-even Higgs decay in the NMSSM */
    struct processes hb; // Second CP-odd Higgs decay in the NMSSM */
    struct processes hc; // Third CP-odd Higgs decay in the NMSSM */
    struct spectrum spectrum; /* summarises the successive annihilations and
                                 decay processes in order to compute the photon 
                                 or pbar spectra */
    double relicmass;
    int NMSSM; /* =0 in the MSSM, =1 in the NMSSM */
};
\end{verbatim}

\subsection{Main routines}

We describe here the main \texttt{SuperIso Relic} routines. The descriptions of the \texttt{SuperIso} and \texttt{AlterBBN} routines are provided in their respective manuals, Refs.~\cite{Mahmoudi:2007vz,Mahmoudi:2008tp,Mahmoudi:2009zz} and Ref.~\cite{Arbey:2011nf,Arbey:2018zfh}.

\subsubsection{SUSY Les Houches Accord routines}

\begin{itemize}
 \item \texttt{int Les\_Houches\_Reader(char name[], struct parameters* param)}\\
 This routine reads the SLHA file \texttt{name} and fills in the \texttt{param} structure.

  \item \texttt{int Higgs\_Decays\_Reader(char name[], struct parameters* param)}\\
\texttt{int SUSY\_Decays\_Reader(char name[], struct parameters* param)}\\
These two routines read the SLHA file \texttt{name} and fill in the Higgs and SUSY decay parts of the \texttt{param} structure.
 
  \item \texttt{int widthcalc(char name[], struct parameters* param)}\\
This routine computes the Higgs decay widths from the SLHA file and fills in the \texttt{param} structure. If \texttt{widthcalc=1} \texttt{Hdecay} is used, otherwise \texttt{FeynHiggs} is called.
  
\end{itemize}

\subsubsection{Relic density}

\begin{itemize}
 \item \texttt{int Weff(double* res, double sqrtS, struct parameters* param)}\\
 This routine computes the effective annihilation rate $W_{\rm eff}$ for a given $\sqrt{S}$ value and puts the results in \texttt{res}. In case of failure, it return 0, otherwise 1.
 
 \item \texttt{void Weff\_table(double Wefftab[][], int *nlines, double sqrtSmax, double maxenergy, struct parameters* param)}\\
 This routine fills in a table of \texttt{nlines} values of the effective annihilation rate $W_{\rm eff}$ for $\sqrt{S}$ between 0 and \texttt{sqrtSmax} or \texttt{maxenergy}. It calls \texttt{Weff} a large number of times. The \texttt{OpenMP} library can be used to call them in parallel.
 
 \item \texttt{double sigmav(double T, double relicmass, double Wefftab[][], int nlines, struct parameters* param)}\\
This routine computes the effective cross section times velocity for a temperature \texttt{T}, using the relic mass \texttt{relicmass} and the $W_{\rm eff}$ table \texttt{Wefftab}.
 
 \item \texttt{void Tfo(double Wefftab[][], int nlines\_Weff, struct parameters* param, double delta, struct relicparam* paramrelic)}\\
This routine determines the freeze-out temperature using the $W_{\rm eff}$ table \texttt{Wefftab}.
 
 \item \texttt{double relic\_density(double Wefftab[][], int nlines\_Weff, struct parameters* param, struct relicparam* paramrelic)}\\
This function returns the relic density using the $W_{\rm eff}$ table \texttt{Wefftab}. This is the central routine which solves the Boltzmann equations.
 
 \item \texttt{double relic\_calculator(char name[])}\\
This routine is a container function which calls different subroutines to obtain the relic density after reading the SLHA file \texttt{name}.
\end{itemize}

For the usage of the other relic density routines, the user can refer to the main programs {\tt test\_phi.c}, {\tt test\_modeleff.c}, {\tt test\_standmod.c}, {\tt test\_reheating.c} and {\tt test\_widthcalc.c}.

\subsubsection{Direct detection}

\begin{itemize}
 \item \texttt{int init\_DDparamSLHA(struct DDparameters* DDparam, char* element, struct parameters* param)}\\
This routine fills in the \texttt{DDparam} structure which is used in the main direct detection routines, using the \texttt{param} parameters. \texttt{element} defines the target material and can be xenon \texttt{"Xe"}, fluorine \texttt{"F"}, germanium \texttt{"Ge"}, argon \texttt{"Ar"}, silicium \texttt{"Si"} or sodium iodide crystals \texttt{"NaI"}.

After this routine is called, the parameters given below can be set to 1 or 0 if one wants to switch on or off the corresponding uncertainties in the calculation of the \texttt{conservative}, \texttt{standard}, and \texttt{stringent} constraints, as defined in Ref.~\cite{Arbey:2017eos}:
\begin{verbatim}
	DDparam.rho0error /* error on the local DM density */
	DDparam.vescerror /* error on the escape velocity */
	DDparam.vroterror /* error on the disk rotational velocity */
	DDparam.nucleonSIerror /* error on the SI nucleon form factors */
	DDparam.nucleonSDerror /* error on the SD nucleon form factors */
	DDparam.nuclearSDerror /* error on the SI nuclear structure factors */
\end{verbatim}	
 
 \item \texttt{void DDsigmaSINucleon(struct DDparameters* DDparam, double* ddpSI, double* ddnSI, char* option)}\\
 \texttt{void DDsigmaSDNucleon(struct DDparameters* DDparam, double* ddpSD, double* ddnSD, char* option)}\\
These two routines compute the spin-independent and spin-dependent scattering cross sections of neutralinos with protons and neutrons from the structure \texttt{DDparam}, and put them in \texttt{ddpSD} and \texttt{ddnSD}. The \texttt{option} can be \texttt{"conservative"}, \texttt{"standard"} or \texttt{"stringent"}, as defined in Ref.~\cite{Arbey:2017eos}.
 
 \item \texttt{double XENON1Tlikelihood(struct DDparameters* DDparam, char *option)}\\
\texttt{double PANDAX2likelihood(struct DDparameters* DDparam, char *option)}\\
\texttt{double PICO60likelihood(struct DDparameters* DDparam, char* option)}\\
These routines compute the likelihoods from the \texttt{DDparam} structure with \texttt{option} \texttt{"conservative"}, \texttt{"standard"} or \texttt{"stringent"} \cite{Arbey:2017eos}, for the experiments XENON1T \cite{Aprile:2017iyp}, PANDAX-2 \cite{Cui:2017nnn} or PICO60 \cite{Amole:2017dex}.

 \item \texttt{double scattering\_SIp\_calculator(char name[])}\\
\texttt{double scattering\_SDp\_calculator(char name[])}\\
These two routines compute the spin-independent and spin-dependent scattering cross sections of the neutralinos with protons, from an SLHA file \texttt{name}.

 \item \texttt{int direct\_xenon1T\_calculator(char name[], char* sigma)}\\
\texttt{int direct\_pandax\_calculator(char name[], char* sigma)}\\
\texttt{int direct\_pico60\_calculator(char name[], char* sigma)}\\
These routines check the exclusion by XENON1T \cite{Aprile:2017iyp}, PANDAX-2 \cite{Cui:2017nnn} or PICO60-60 \cite{Amole:2017dex} from an SLHA file \texttt{name} at a given confidence level such as \texttt{"2sigma"}. They return 1 if the model point is excluded and 0 if it is not.

 \item \texttt{void dRdE(struct DDparameters* DDparam, int n,double E[], double dRdE0[], char* option)}\\
This routine calculates the differential recoil rate per unit of target material mass for the particle model and target material defined in \texttt{DDparam} and for \texttt{n} recoil energy values in array \texttt{E}. The result is stored in \texttt{dRdE0}. \texttt{option} can be \texttt{conservative}, \texttt{standard} or \texttt{stringent}.

\item \texttt{void set\_nucleonFFSI(struct DDparameters* DDparam, double sigmapin[2],double sigmas[2], double z[2])}\\
This routine can be used to modify the main values and errors on the parameters $\Sigma_{\pi N}$, $\sigma_s$ and $z$ used in the calculation of SI nucleon form factors. The first entry of \texttt{sigmapin}, \texttt{sigmas} and \texttt{z} must contain the central value and the second entry the error. For instance, $\Sigma_{\pi N}= \texttt{sigmapin[0]} \pm \texttt{sigmapin[1]}$.

\item \texttt{void set\_nucleonFFSD(struct DDparameters* DDparam,double a3[2], double a8[2], double fsSDp[2])}\\
Similarly to \texttt{set\_nucleonFFSI}, this routine allows the user to modify the values of parameters $a_3$, $a_8$ and $\Delta s^{p}$ used in the calculation of SD nucleon form factors.

\item \texttt{void set\_nuclei\_SIformfactor(double (*FFSI)(double,int))}\\
This routine allows the user to define another SI nuclear form factor \texttt{FFSI} rather than the default Helm function. \texttt{FFSI} takes as input the momentum transfer $q$ in keV and the nucleon number of the nucleus $A$.

\item \texttt{void set\_nuclei\_SDformfactornoerr(struct DDparameters* DDparam,double (*S00[]) (double),double (*S01[])(double),double (*S11[])(double))}\\
This routine allows the user to define alternative SD structure factors $S_{00}$, $S_{01}$, and $S_{11}$. The inputs are arrays containing the structure factors of each isotope.

\item \texttt{void set\_Maxwell(struct DDparameters* DDparam, double rho0[], double vrot[], double vearth[], double vesc[])}\\
This routine allows the user to define alternative parameter values for the standard halo model. Each input contains the main value and error of the parameter. \texttt{rho0} is the local DM density, \texttt{vrot} is the galactic disk rotation velocity, \texttt{vearth} is the Earth velocity in the local standard of rest and \texttt{vesc} is the escape velocity.

\end{itemize}

For the usage of the other direct detection routines, the user can read the main program {\tt direct.c}.

\subsubsection{Indirect detection}
\paragraph{Standard routines}
\begin{itemize}
 \item \texttt{int init\_IDparameters(struct parameters* param, struct IDparameters* IDparam)}\\
This routine fills in the \texttt{IDparam} structure which is used in the main indirect detection routines, using the \texttt{param} parameters.

 \item \texttt{double deltalikelihood\_fermi(struct fermi* fe, struct IDparameters* IDparam, char* option)}\\
 This routine computes the likelihood from the \texttt{IDparam} structure with \texttt{option} \texttt{"conservative"}, \texttt{"standard"} or \texttt{"stringent"} \cite{Arbey:2017eos} for the FERMI-LAT experiment \cite{Ackermann:2015zua}.

 \item \texttt{double deltachi2\_AMS(struct IDparameters* IDparam, char* option)}\\
 This routine computes the $\Delta \chi^2$ from the \texttt{IDparam} structure with \texttt{option} \texttt{"conservative"}, \texttt{"standard"} or \texttt{"stringent"} \cite{Arbey:2017eos} for the AMS-02 experiment \cite{Aguilar:2016kjl}.

 \item \texttt{double annihilation\_sigmav\_calculator(char name[])}\\
 This routine computes the total annihilation cross section for a given SLHA file \texttt{name}.

 \item \texttt{int indirect\_fermi\_calculator(char name[], char*sigma)}\\
\texttt{int indirect\_ams02\_calculator(char name[], char*sigma)}\\
These routines check the exclusion by FERMI-LAT \cite{Ackermann:2015zua} or AMS-02 \cite{Aguilar:2016kjl} from an SLHA file \texttt{name} at a given confidence level such as \texttt{"2sigma"}. They return 1 if the model point is excluded and 0 if it is not.

\end{itemize}
For the usage of the other indirect detection routines, the user can read the main program {\tt indirect.c}.

\paragraph{Propagation and dark matter halo models}

\begin{itemize}
 \item \texttt{void init\_mypropagation(struct propagation\_parameters* pparam, double DIFFUSION\_0\_GV, double PUISSANCE\_COEFF\_DIFF, double E\_DIFFUS, double VENT\_GALACTIQUE, double V\_ALFEN, char* propname)}\\
This routine is used in the context of indirect detection to specify a specific antiproton propagation model. The ordered input parameters are the coefficients $\mathcal{K}_0$, $\delta$, $L$, $V_{conv}$ and $V_a$, which are defined in Appendix \ref{app:indirect}.

 An example of its usage is given in the main program \texttt{create\_propagation.c}.
 
 \item \texttt{void init\_myhalo(struct propagation\_parameters* pparam, double r\_earth, double rho\_chi\_solar, double (*profile)(double,double), char* name)}\\
This routine is used in the context of indirect detection to specify a dark matter halo profile. An example of its usage is given in the main program \texttt{create\_propagation.c}. \texttt{r\_earth} is the distance of Earth from the galactic center in kpc, \texttt{rho\_chi\_solar} the local DM density in GeV/cm$^3$, \texttt{profile} is a function giving the DM density as a function of the distance from the galactic center $r$ and the height $z$ from the galactic plane.

\item \texttt{void init\_existing\_model(struct propagation\_parameters* pparam, char * halo, char* propagation)}\\
This routine is used to initialise a dark matter halo profile and the propagation model already provided in the package. \texttt{halo} can be \texttt{Einasto\_CU10}, \texttt{NFW\_M16} or \texttt{Burkert\_NS13} and \texttt{propagation} can be \texttt{MIN}, \texttt{MED} or \texttt{MAX}.

\item \texttt{void write\_secondaries(struct propagation\_parameters* pparam)}\\
This routine is used to write tabulated secondary antiproton spectra for the propagation model specified in \texttt{pparam}. The result can be found in the directory\\
 \texttt{src/antiprotons/models/haloname/propname/secondaries/} , with \texttt{haloname} and \texttt{propname} the names of the halo model and propagation model specified in \texttt{pparam}.

\item \texttt{void write\_primaries(struct propagation\_parameters* pparam, double mass\_inf, double mass\_sup)}\\
Similarly to \texttt{write\_secondaries}, this routine writes primary antiproton spectra in the directory \texttt{src/antiprotons/models/haloname/propname/primaries/} for LSP masses from \texttt{mass\_inf} to \texttt{mass\_sup}. \texttt{mass\_inf} must be greater than 5 GeV and \texttt{mass\_inf} must be smaller than 100000 GeV. Once the tabulated secondary and primary spectra are written, the calculation of AMS-02 antiproton constraints is much faster, which can be useful when performing large scans.

\end{itemize}

For further descriptions of the propagation routines, the user can refer to the files {\tt src/propagation.c} and {\tt src/propagation.h}.
\subsubsection{Big-Bang Nucleosynthesis}

\begin{itemize}
\item \texttt{int bbn\_excluded(struct relicparam* paramrelic)}\\
This routine belongs to \texttt{AlterBBN}. It returns 1 if the cosmological scenario defined in the \texttt{paramrelic} structure is excluded by the Big-Bang nucleosynthesis constraints, 0 if not, and -1 if the routine fails.
\end{itemize}

For the usage of the \texttt{AlterBBN} routines, the user can refer to the \texttt{AlterBBN} manual \cite{Arbey:2011nf,Arbey:2018zfh}.

\section{Compilation and installation instructions}%
\label{compilation}
The \texttt{SuperIso Relic} package and the latest version of the manual can be downloaded from:
\begin{center}
 \url{http://superiso.in2p3.fr/relic}
\end{center}
The package can be uncompressed with
\begin{center}
 \tt tar xjvf superiso\_relic\_vX.X.tar.bz2
\end{center}
Before configuration, the user can choose the compilers with \texttt{BASH} commands such as\\
{\tt export CC=gcc\\
export FC=gfortran\\
export CXX=g++}\\
The automatic configuration can be run with for example
\begin{center}
 \tt ./configure --with-mp
\end{center}
\texttt{"--with-mp"} is an option to activate the \texttt{OpenMP} multiprocessor compilation and calculation. The \texttt{configure} options can be accessed with \texttt{configure --help}.

At this step, it is possible to modify several configuration options in \texttt{Makefile}:
\begin{itemize}
 \item \texttt{RELIC} can be set to 0 to deactivate the dark matter routines (this would correspond to the usage of \texttt{SuperIso} only for the calculation of the flavour observables), 1 to activate them only for the MSSM, and 2 for the MSSM and NMSSM. The main consequences of this choice are the compilation time and the size of the compiled library.
 \item The links to \texttt{SOFTSUSY}, \texttt{ISAJET}, \texttt{SuSpect}, \texttt{SPheno} and \texttt{NMSSMTools} have to be given here.
 \item The multiprocessor configuration can be forced by uncommenting \texttt{CFLAGS\_MP}, \texttt{FFLAGS\_MP} and \texttt{MAKE\_MP}.
\end{itemize}

\noindent The next step is to configure the compilation mode with
\begin{center}
 \tt make shared
\end{center}
which compiles the dark matter Fortran routines in shared dynamic libraries, or
\begin{center}
 \tt make static
\end{center}
which compiles the routines in static libraries. The shared mode is faster to compile, but slower at execution, since the Fortran routines are compiled on-the-fly. The static mode is slower to compile, but easy to export.

The compilation of the libraries starts with
\begin{center}
 \tt make
\end{center}
and the main programs are compiled with
\begin{center}
 \tt make main\_program.c
\end{center}

Examples of executions of main programs are given in the next section. In the static mode, if the \texttt{OpenMP} library is used, it is possible that a segmentation fault appears at execution, because of the large number of \texttt{COMMON} blocks used in the Fortran routines. In such a case, the \texttt{BASH} command
\begin{center}
 \tt export OMP\_STACKSIZE=16M
\end{center}
should be run before the execution.

%
\section{Input and output description}%
\label{sample}

\subsection{Standard cosmological and astrophysical scenarios}

The program \texttt{slha.x} calculates the observables using the parameters contained in a given SLHA file. The calculations are performed in the standard cosmological and astrophysical scenarios. For example, the command
\begin{verbatim}
./slha.x example.lha 
\end{verbatim}
returns
\begin{verbatim}
SuperIso Relic v4.0 - A. Arbey, F. Mahmoudi & G. Robbins
SuperIso v4.0 - F. Mahmoudi

Observable                      Value

BR(b->s gamma)                  3.644e-04
delta0(B->K* gamma)             4.307e-02
\end{verbatim}
\[
\huge \bf \vdots
\]
\begin{verbatim}
Relic density Oh2               1.254e+01

SI proton xsection              1.839e-10
SD proton xsection              5.242e-07

excluded_Xenon1T (standard)     0
excluded_PANDAX (standard)      0
excluded_PICO60 (standard)      0

Tot annihilation xsection       4.084e-30

excluded_Fermi (standard)       0
excluded_AMS02 (standard)       0
\end{verbatim}
here 0 means that the point is not excluded by the given constraint.

\subsection{Modified relic density}

\texttt{SuperIso Relic} incorporates different cosmological scenarios which can affect the relic density calculation. The description of the scenarios is provided in the appendices, and their usage is explained below.

\subsubsection{Alternative QCD equations of state}
The program \texttt{test\_modeleff.x} calculates the relic density using the parameters contained in a given SLHA file, for different QCD equations of state. For example, the command
\begin{verbatim}
./test_modeleff.x example.lha
\end{verbatim}
returns
\begin{verbatim}
Dependence of the relic density on the calculation of heff and geff
For model_eff=1 (model A): omega=1.254e+01
For model_eff=2 (model B (default)): omega=1.254e+01
For model_eff=3 (model B2): omega=1.262e+01
For model_eff=4 (model B3): omega=1.247e+01
For model_eff=5 (model C): omega=1.255e+01
For model_eff=6 (Bonn model): omega=1.231e+01
For model_eff=0 (old model): omega=1.229e+01
\end{verbatim}
The description of the different models are provided in Appendix \ref{qcdstate}.

\subsubsection{Effective energy and entropy densities}

The program \texttt{test\_standmod.x} calculates the relic density using the parameters contained in a given SLHA file, in a cosmological scenario where the energy and entropy content is modified. In addition to the name of an SLHA file, this program needs four additional parameters and five optional ones, and is run with:
\begin{center}
\texttt{./test\_standmod.x} filename $\kappa_\rho$ $n_\rho$ $\kappa_s$ $n_s$ ($T_\rho$ $T_s$ $N$ $n_N$ $T_N$)
\end{center}
where
\begin{itemize}
 \item $\kappa_\rho$ is the ratio of dark energy density to photon energy density at $T_{\rm BBN} = 1$ MeV,
 \item $n_\rho$ is the temperature exponent of the dark energy density,
 \item $\kappa_s$ is the ratio of dark entropy density to photon entropy density at $T_{\rm BBN} = 1$ MeV,
 \item $n_s$ is the temperature exponent of the dark entropy density,
\end{itemize}
and the optional parameters (which, if not specified, are set to 0) are:
\begin{itemize}
 \item $T_\rho$ is a temperature cut in GeV below which the dark energy density is set to 0,
 \item $T_s$ is a temperature cut in GeV below which the dark entropy density is set to 0,
 \item $N$ is the non-thermal production rate of supersymmetric particles at $T_{\rm BBN} = 1$ MeV,
 \item $n_N$ is the temperature exponent of the non-thermal production rate of supersymmetric particles,
 \item $T_N$ is a temperature cut in GeV below which the non-thermal production is set to 0,
\end{itemize}
which are further described in Appendix~\ref{mocomo}.

For example, the command
\begin{verbatim}
./test_standmod.x example.lha 1 6 1 5
\end{verbatim}
returns
\begin{verbatim}
SuperIso Relic v4.0 - A. Arbey, F. Mahmoudi & G. Robbins
AlterBBN v2.0 - A. Arbey, J. Auffinger, K. Hickerson & E. Jenssen

For the cosmological standard model:
omega=1.254e+01
For the specified model with dark density/entropy/non thermal relics:
omega=9.924e+03
Model excluded by BBN constraints
\end{verbatim}

\subsubsection{Reheating}

The program \texttt{test\_reheating.x} calculates the relic density using the parameters contained in a given SLHA file, in a cosmological scenario with reheating. In addition to the name of an SLHA file, this program needs five additional parameters and six optional ones, and is run with:
\begin{center}
\texttt{./test\_reheating.x} filename $\kappa_\rho$ $n_\rho$ $\kappa_{\Sigma_r}$ $n_{\Sigma_r}$ $T_{\rm cut}$ ($\kappa_s$ $n_s$ $N$ $n_N$ $\kappa_\Sigma$ $n_\Sigma$)
\end{center}
\begin{itemize}
 \item $\kappa_\rho$ is the ratio of dark energy density to photon energy density at $T_{\rm BBN} = 1$ MeV,
 \item $n_\rho$ is the temperature exponent of the dark energy density,
 \item $\kappa_{\Sigma_r}$ is the ratio of radiation entropy injection $\Sigma_r$ to radiation entropy time-derivative at $T_{\rm BBN} = 1$ MeV,
 \item $n_{\Sigma_r}$ is the temperature exponent of the radiation entropy injection,
 \item $T_{\rm cut}$ is a temperature cut in GeV below which non-standard terms are set to 0,
\end{itemize}
and the optional parameters (which, if not specified, are set to 0) are:
\begin{itemize}
 \item $\kappa_s$ is the ratio of dark entropy density to photon entropy density at $T_{\rm BBN} = 1$ MeV,
 \item $n_s$ is the temperature exponent of the dark entropy density,
 \item $N$ is a non-thermal production rate of supersymmetric particles at $T_{\rm BBN} = 1$ MeV,
 \item $n_N$ is the temperature exponent of the non-thermal production rate of supersymmetric particles,
 \item $\kappa_\Sigma$ is the ratio of dark entropy injection $\Sigma$ to radiation entropy time-derivative at $T_{\rm BBN} = 1$ MeV,
 \item $n_\Sigma$ is the temperature exponent of the dark entropy injection,
\end{itemize}
which are further described in Appendix~\ref{mocomo}.

For example, the command
\begin{verbatim}
./test_reheating.x example.lha 0 0 1.5 5 1e-2
\end{verbatim}
returns
\begin{verbatim}
For the cosmological standard model:
omega=1.254e+01
For the specified model with dark density/radiation entropy production:
omega=4.854e+00
Model compatible with BBN constraints
\end{verbatim}

\subsubsection{Decaying scalar field}

The program \texttt{test\_phi.x} calculates the relic density using the parameters contained in a given SLHA file, in a cosmological scenario with a decaying scalar field. In addition to the name of an SLHA file, this program needs two additional parameters and two optional ones, and is run with:
\begin{center}
\texttt{./test\_phi.x} filename $\tilde\rho_\phi$ $T_{RH}$ ($T_{\rm init}$ $\eta_\phi$)
\end{center}
where
\begin{itemize}
 \item $\tilde\rho_\phi$ is the ratio of the scalar field density to photon energy density at $T=T_{\rm init}$.
 \item $T_{RH}$ is the reheating temperature in GeV,
\end{itemize}
and two optional ones:
\begin{itemize}
 \item $T_{\rm init}$ is the initial temperature in GeV at which the calculation starts (100 GeV by default),
 \item $\eta_\phi$ is the non-thermal supersymmetric particle production parameter,
\end{itemize}
which are further described in Appendix~\ref{app:scalar}.

For example, the command
\begin{verbatim}
./test_phi.x example.lha 100 10 50
\end{verbatim}
returns
\begin{verbatim}
For the cosmological standard model:
omega=1.220e+01
For the specified model with dark density/entropy/non thermal relics:
omega=3.989e+00
Model compatible with BBN constraints
\end{verbatim}
Note that for this scenario \texttt{full\_comput} is set to 1 in order to compute the evolution of the scalar field before freeze-out. The calculation is therefore slower and the needed time increases with the initial temperature.

\subsubsection{Width calculators}

The program \texttt{test\_widthcalc.x} calculates the relic density using the parameters contained in a given SLHA file, and the Higgs widths computed with different decay calculators. For example, the command
\begin{verbatim}
./test_widthcalc.x example.lha
\end{verbatim}
returns
\begin{verbatim}
Dependence of the relic density on the width calculator
Widths in the SLHA file: omega=1.254e+01
With Hdecay: omega=1.254e+01
With FeynHiggs: omega=1.254e+01
With FeynHiggs Tree: omega=1.254e+01
\end{verbatim}
which shows that for this specific point, the dependence on the Higgs widths is negligible.

\subsection{Direct detection}

The program \texttt{direct.x} calculates dark matter direct detection observables using the parameters contained in a given SLHA file. In addition to the name of an SLHA file, this program needs two additional parameters \texttt{nuclerr} and \texttt{haloerr}, which if set to 1 activate the uncertainties related to the nuclear form factors and halo profile, and if 0 deactivate them. For example, the command
\begin{verbatim}
./direct.x example.lha 1 1
\end{verbatim}
returns
\begin{verbatim}
/----WIMP-NUCLEON cross-section(pb) at 0-momentum transfer-----/

 /-Spin-Independent-/
Conservative             Standard       Stringent
Proton 1.560368e-10     1.838746e-10    2.166668e-10
Neutron 1.601246e-10    1.907715e-10    2.242631e-10

/-Spin-Dependent-/
Conservative             Standard       Stringent
Proton 4.702998e-07     5.242344e-07    5.810959e-07
Neutron 4.788661e-07    5.332762e-07    5.906135e-07

/------PANDAX-2 2017 likelihood (point excluded at 90% C.L.if >2.710000)------/
Conservative     Standard        Stringent
 8.611648e-02    2.265334e+00     6.704124e+00

 
/-------XENON1T 2017 likelihood (point excluded at 90% C.L if >2.710000)------/
Conservative     Standard        Stringent
 8.199753e-01    2.076444e+00     3.857909e+00

/----PICO60 2017 likelihood (point excluded at 90% C.L if >2.710000)------/
Conservative     Standard        Stringent
 7.838017e-03    2.261627e-02     4.457053e-02
\end{verbatim}


\subsection{Indirect detection}

\subsubsection{Standard scenarios}

The program \texttt{indirect.x} calculates dark matter indirect detection observables using the parameters contained in a given SLHA file, for standard astrophysical scenarios. For example, the command
\begin{verbatim}
./indirect.x example.lha 
\end{verbatim}
returns
\begin{verbatim}
neutralino mass 209.466537 GeV

channel           <sigmav>[cm^3/s]
---------------------------------
o1 o1 -> ebar e |   6.50906e-38
o1 o1 -> mbar m |   2.78280e-33
o1 o1 -> lbar l |   5.06090e-31
o1 o1 -> ubar u |   8.05216e-39
o1 o1 -> dbar d |   7.72181e-36
o1 o1 -> sbar s |   3.22156e-33
o1 o1 -> cbar c |   5.86786e-33
o1 o1 -> bbar b |   2.87770e-30
o1 o1 -> tbar t |   1.73053e-31
o1 o1 -> wbar w |   2.33753e-31
o1 o1 -> z z    |   7.59111e-32
o1 o1 -> h z    |   2.06015e-31
---------------------------------
Total annihilation cross section 4.084403e-30 [cm^3/s]

---------------------------------
Fermi-LAT
---------------------------------
delta-loglikelihood : excluded at 2 sigma if <-6.425000
---------------------------------
conservative     standard        stringent
---------------------------------
3.242732e-06     3.063343e-06    -2.679171e-06

---------------------------------
AMS-02
---------------------------------
delta-chi^2 : excluded at 2 sigma if >12.850000
---------------------------------
conservative     standard        stringent
---------------------------------
-1.658848e-02    -2.208422e-02   -7.297001e-02
\end{verbatim}

\subsubsection{Scenarios with modified propagation and halo models}

The program \texttt{create\_propagation.x} calculates dark matter indirect detection observables using the parameters contained in a given SLHA file, for the dark matter halo and propagation scenarios defined in \texttt{create\_propagation.c}. For example, the command
\begin{verbatim}
./create_propagation.x example.lha 
\end{verbatim}
returns
\begin{verbatim}
neutralino mass 209.466537 GeV

channel           <sigmav>[cm^3/s]
---------------------------------
o1 o1 -> ebar e |   6.50906e-38
o1 o1 -> mbar m |   2.78280e-33
o1 o1 -> lbar l |   5.06090e-31
o1 o1 -> ubar u |   8.05216e-39
o1 o1 -> dbar d |   7.72181e-36
o1 o1 -> sbar s |   3.22156e-33
o1 o1 -> cbar c |   5.86786e-33
o1 o1 -> bbar b |   2.87770e-30
o1 o1 -> tbar t |   1.73053e-31
o1 o1 -> wbar w |   2.33753e-31
o1 o1 -> z z    |   7.59111e-32
o1 o1 -> h z    |   2.06015e-31
---------------------------------
Total annihilation cross section 4.084403e-30 [cm^3/s]
dchi2 -4.444347e-02
\end{verbatim}

\section{Results}
\label{result}
\begin{figure}
\centering
\begin{minipage}[t]{0.46\linewidth}
\includegraphics[width=\textwidth]{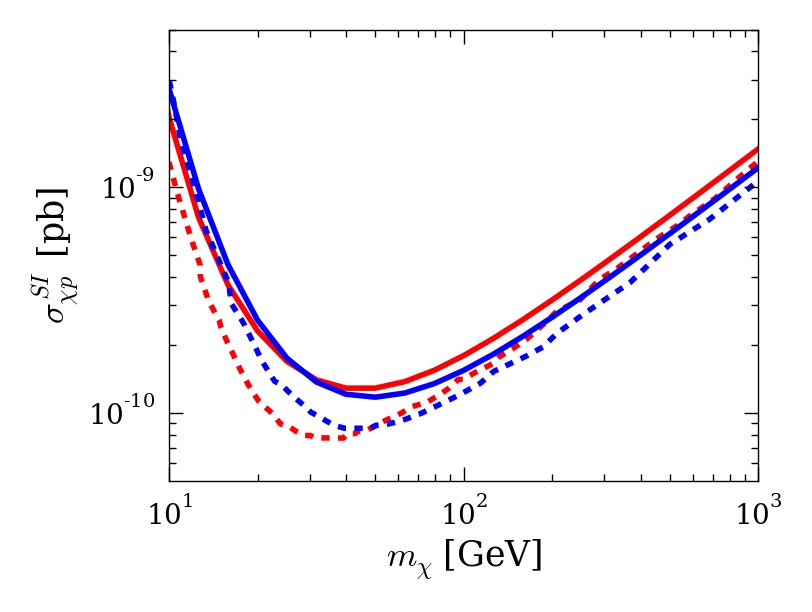}
\caption[90\% C.L. spin-independent upper limit for XENON1T, PANDAX-2 and DARWIN]{90\% C.L. spin-independent upper limit for XENON1T (red) and PANDAX-2 (blue). Official limits published by the collaborations are shown in dashed lines while the limits calculated from this work are in solid lines.}
\label{fig:limSI}
\end{minipage}
\quad
\begin{minipage}[t]{0.46\linewidth}
\centering
\includegraphics[width=\textwidth]{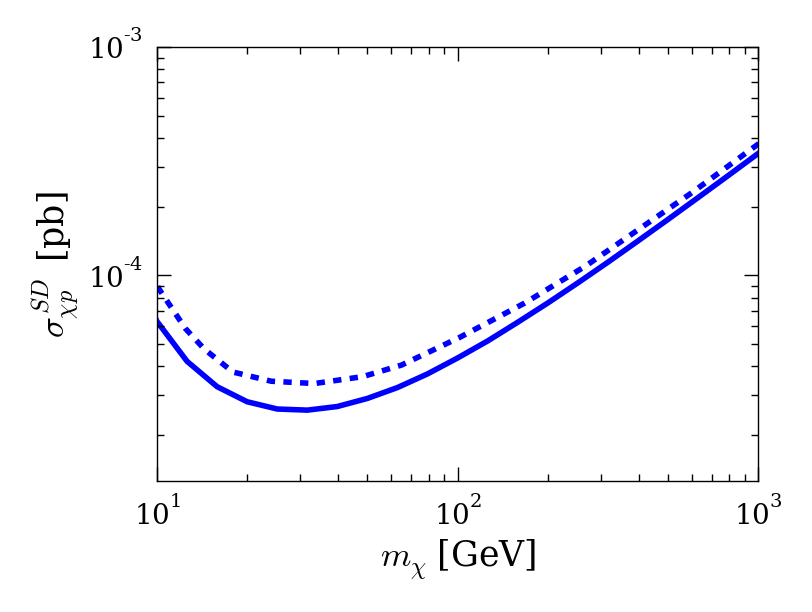}
\caption[90\% C.L. spin-dependent upper limit for PICO60]{90\% C.L. spin-dependent upper limit for PICO60. Official limit published by the collaboration is shown in dashed line while the limit calculated from this work is in solid line.}
\label{fig:limSD}
\end{minipage}
\end{figure}

The results of \texttt{SuperIso Relic} have been extensively compared with those of \texttt{micrOMEGAs} \cite{Belanger:2006is,Belanger:2008sj,Belanger:2013oya} and \texttt{DARKSUSY} \cite{Gondolo:2004sc}, and show a very good global agreement.\\
The implementation of indirect and direct detection constraints is detailed in Appendices \ref{app:indirect} and \ref{app:direct} and we show here the spin-independent (SI) upper limits from XENON1T \cite{Aprile:2017iyp} and PANDAX-2 \cite{Cui:2017nnn}, as well as the spin-dependent (SD) limit from PICO60 \cite{Amole:2017dex} in figures \ref{fig:limSI} and \ref{fig:limSD}, respectively. The official limits are also displayed for comparison. For the calculation of the upper limits, we used the standard parameter values of the dark matter halo $\rho_{\chi}=0.3$ GeV/cm$^3$, $v_{rot}=220$ km/s and $v_{esc}=544 $ km/s.

\begin{figure}[t!]
\centering
\begin{minipage}[t]{0.46\linewidth}
\centering
\includegraphics[width=\textwidth]{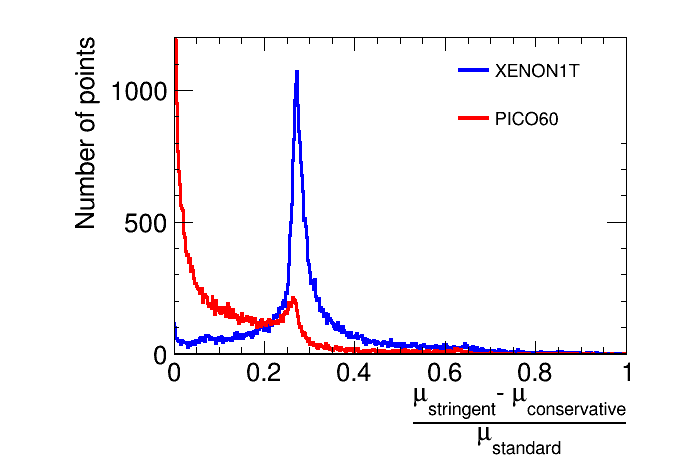}
\caption{Relative uncertainties from SI nucleon form factor errors on the number of events which would be measured by XENON1T (in blue) and PICO60 (in red) for our pMSSM sample of points.}
\label{fig:nucleonSI}
\end{minipage}
\quad
\begin{minipage}[t]{0.46\linewidth}
\centering
\includegraphics[width=\textwidth]{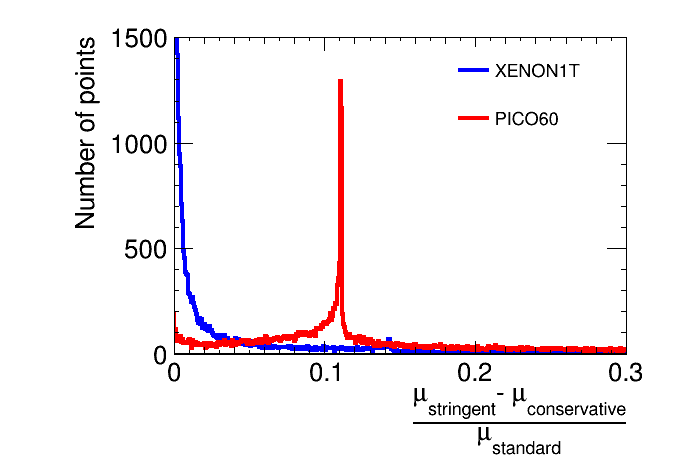}
\caption{Relative uncertainties from SD nucleon form factor errors on the number of events which would be measured by XENON1T (in blue) and PICO60 (in red) for our pMSSM sample of points.}
\label{fig:nucleonSD}
\end{minipage}
\end{figure}

In addition, we show the impact of nuclear errors on these limits and thus complete the panel of uncertainties studied in \cite{Arbey:2017eos}. To this end, we use the same pMSSM-19 sample of points as in Ref.~\cite{Arbey:2017eos} and calculate for each point the number of expected events $\mu$ which would be measured by XENON1T and PICO60 in the \enquote{conservative}, \enquote{standard} and \enquote{stringent} case, according to the underlying uncertainties, and show the relative error $(\mu_{\rm stringent} - \mu_{\rm conservative})/\mu_{\rm standard}$. The relative error coming from SI nucleon form factors is first shown in figure \ref{fig:nucleonSI}. One can note a typical error of $\sim 27$\% for the xenon experiment, whereas the error can be neglected in most cases for the fluorine experiment, which is more sensitive to SD interactions. By default, we will therefore consider the SI nucleon form factor uncertainties for xenon experiments but not for fluorines experiments. Then, we show in figure \ref{fig:nucleonSD} the relative error from SD nucleon form factor uncertainties. For PICO60, we obtain a typical error of $\sim 11$\%, whereas the uncertainties can be safely neglected for XENON1T. In this last case, SD nucleon form factor uncertainties are therefore, by default, not considered. 

\begin{figure}[t!]
\centering
\includegraphics[width=0.6\textwidth]{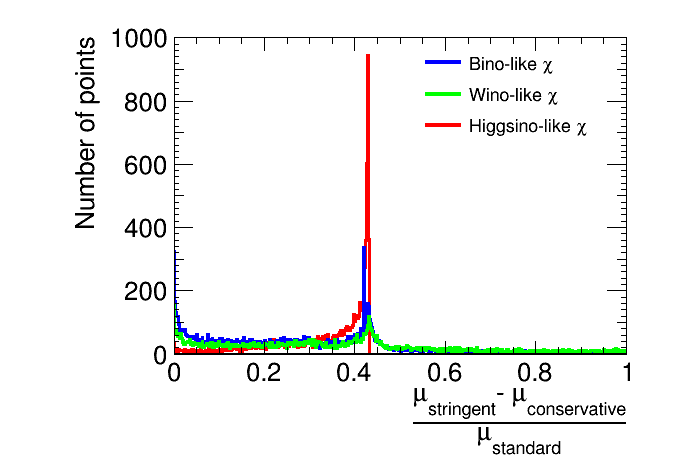}
\caption{Relative uncertainties from SD nuclear structure factor errors on the number of events which would be measured by PICO60 for our pMSSM sample of points. Wino-like neutralinos are shown in green, Higgsino-like neutralinos in red and bino-like neutralinos in blue.}
\label{fig:nuclearSD}
\end{figure}

Finally, we show in figure \ref{fig:nuclearSD}, the impact of the uncertainties on the SD nuclear structure factors arising from two-body currents. While the effect on XENON1T results can be safely disregarded, the relative error on PICO60 detection is of the order of 40\%. Higgsino-like neutralinos suffer in particular from these uncertainties. Indeed, in the structure factor $S(q)$, the functions $S_{01}$ and $S_{11}$, which have the uncertainties, are multiplied by $a_1=A_p^{SD}-A_n^{SD}$ (see appendix \ref{app:direct}). Spin-dependent nuclear uncertainties therefore have an effect for isospin violating models. As the effective SD neutralino-quark coupling through a $Z$ boson exchange is proportional to the quark weak isospin $t^3$ and to the difference of the Higgsino mixing matrix elements $|N_{1\,4}|^2 - |N_{1\,3}|^2$, Higgsino-like neutralinos have thereby a particularly large $a_1$ and are therefore strongly affected by the structure function uncertainties.

%
\newpage
\appendix
%
\append{Relic density in alternative cosmology}

We provide here the description of the modifications to the cosmological scenarios focussing in particular on the new aspects with respect to the original manual \cite{Arbey:2009gu,Arbey:2011zz}.

\subsection{QCD equations of state}
\label{qcdstate}
The number of relativistic effective radiation degrees of freedom of energy and entropy, $g_{\mbox{eff}}$ and $h_{\mbox{eff}}$ respectively, are important for the calculation of the relic density. The previous versions of \texttt{SuperIso Relic} incorporated the ideal gas scenarios as well as five more realistic models described in Ref.~\cite{Hindmarsh:2005ix}. The current version incorporates in addition another scenario for which lattice QCD calculations are used to improve the description of the thermal plasma, which is precisely described in Ref. \cite{Drees:2015exa}, and which is referred to as \texttt{Bonn model} in \texttt{SuperIso Relic}. 

The models can be selected in the routine \texttt{Init\_modeleff(int model\_eff, struct relicparam* paramrelic)} by setting the value of \texttt{model\_eff} as given below:
\begin{itemize}
\item Model A (\verb?model_eff?=1): ignores hadrons completely.
\item Model B (\verb?model_eff?=2): models hadrons as a gas of free mesons and hadrons, with a sharp switch to the hadronic gas at a temperature $T = 154$ MeV.
\item Model B2 (\verb?model_eff?=3): variation of model B constructed by scaling the pressure and energy density lattice data by 0.9.
\item Model B3 (\verb?model_eff?=4): variation of model B constructed by scaling the pressure and energy density lattice data by 1.1.
\item Model C (\verb?model_eff?=5): models hadrons as a gas of free mesons and hadrons, with a sharp switch to the hadronic gas at $T = 200$ MeV.
\item Bonn model (\verb?model_eff?=6): uses lattice QCD calculations to improve the description of the thermal plasma.
\item Old Model (\verb?model_eff?=0): models hadrons as an ideal gas.
\end{itemize}
An example main program is given in \texttt{test\_modeleff.c}.

\subsection{Modified Cosmological Models}%
\label{mocomo}

The number density of supersymmetric particles can be determined by solving the Boltzmann equation:
\begin{equation}
\frac{dn}{dt} = - 3 H n - \langle \sigma v \rangle (n^2 - n^2_{eq}) + N_{n.t.}  \,,\label{boltzmann}
\end{equation}
where $n$ is the number density of supersymmetric particles, $\langle \sigma v \rangle$ the thermally averaged annihilation cross-section, $H=\dot{a}/a$ the Hubble parameter, $a(t)$ the expansion factor, and $n_{eq}$ the relic particle equilibrium number density. The term $N_{n.t.}$ corresponds to a possible non-thermal production of SUSY particles. The expansion rate $H$ is related to the energy content of the Universe through the Friedmann equation:
\begin{equation}
 H^2=\frac{8 \pi G}{3} (\rho_{rad} + \rho_D)  \,,\label{friedmann}
\end{equation}
where $\rho_{rad}$ is the radiation energy density, which is considered as dominant before BBN in the standard cosmological model. The radiation energy density reads:
\begin{equation}
\rho_{rad}=g_{\rm{eff}}(T) \frac{\pi^2}{30} T^4\,.
\end{equation}
$\rho_D$ can be considered as an effective ``dark energy'' density which parametrises the modification of the expansion rate.

The relation between the time $t$ and/or scale factor $a(t)$ and the temperature $T$ is given by the entropy evolution equation:
\begin{equation}
\frac{ds}{dt} = - 3 H s + \Sigma \label{entropy_evolution} \,,
\end{equation}
where $s$ is the entropy density and $\Sigma$ an entropy injection. In the standard model of cosmology $s$ contains the radiation entropy density only, which depends on the QCD equations of state, and $\Sigma=0$. The radiation entropy density can be written as:
\begin{equation}
s_{rad} = h_{\rm{eff}}(T) \frac{2\pi^2}{45} T^3 \,.
\end{equation}

In a modified cosmological scenario, the entropy density can be composed of radiation entropy density and a ``dark entropy'' density $s_D$. $\Sigma$ can therefore be an injection term either for radiation entropy or dark entropy.

In the following, we describe the different parametrisations implemented in \texttt{SuperIso Relic}.

\subsubsection{Modified expansion rate}

The expansion rate is modified through the addition of a ``dark density'' $\rho_D$, which is parametrised as \cite{Arbey:2008kv}:
\begin{equation}
\rho_D(T) = \kappa_\rho \, \rho_{\gamma}(T_{\rm BBN})\left(\frac{T}{T_{\rm BBN}}\right)^{n_\rho}\,,
\end{equation}
where $T_{\rm BBN} = 1$ MeV is close to the Big-Bang Nucleosynthesis scale. $\kappa_\rho$ is therefore the ratio of the dark density to the photon density at this energy scale\footnote{In the first versions of \texttt{SuperIso Relic}, $\kappa_\rho$ was defined as the ratio of the dark density to the \emph{radiation} density.}. $n_\rho$ is the decrease exponent of this dark density (4 for radiation, 3 for matter, 0 for a cosmological constant, \textit{etc.}). A temperature cut can be added below which the dark density is set to 0.

\subsubsection{Modification of the entropy content}

We consider two different cases for the entropy content, which can occur simultaneously or separately.

\paragraph{1 -- Dark entropy:}
If the ``dark entropy'' $s_D$ is not linked to any reheating process of radiation, two different parametrisations are possible. The first one was described in Refs.~\cite{Arbey:2008kv,Arbey:2009gt}:
\begin{equation}
	s_D = \kappa_s \, s_\gamma(T_{\rm BBN}) \left( \frac{T}{T_{\rm BBN}} \right)^{n_s}\,,
\end{equation}
where $\kappa_s$ is the ratio of the dark entropy density to the photon entropy density at $T_{\rm BBN} = 1$ MeV, and $n_s$ is the decrease exponent of this component. Again, a temperature cut below which this density is strictly 0 can be set. 

The second parametrisation of the ``dark entropy'' density is through the dark entropy injection $\Sigma_D$ as described in Ref.~\cite{Arbey:2011gu}:
\begin{equation}
	\Sigma_D(T) = \kappa_\Sigma \, \Sigma_{rad}^{\rm eff}(T_{\rm BBN})\left( \frac{T}{T_{\rm BBN}} \right)^{n_\Sigma}\,,
\end{equation}
where $\kappa_\Sigma$ is ratio of the dark entropy injection to the time-variation of the radiation entropy density time-derivative $\Sigma_{rad}^{\rm eff}(T) \equiv \left| \dfrac{ds_{rad}}{dt}\right| = 3 H s_{rad}$ at $T_{\rm BBN} = 1\,$MeV and $n_\Sigma$ is the decrease exponent of this production. The associated ``dark entropy'' density is therefore the integral:
\begin{equation}
	s_D(T) = 3\sqrt{\frac{5}{4\pi^3 G}} h_{\rm{eff}}(T) T^3 \int_0^T {\rm d}T^\prime \frac{ \sqrt{g_*(T)} \Sigma_D(T^\prime)}{h_{\rm{eff}}^2(T^\prime)T^{\prime\,6}\sqrt{1+\rho_D(T)/\rho_{rad}(T)}}\,,
\end{equation}
where the values of $h_{\rm{eff}}(T)$ and $g_{\rm{eff}}(T)$ are tabulated.

\paragraph{2 -- Reheating: }
The radiation entropy density $s_{rad}$ can receive contributions from a radiation entropy injection $\Sigma_{rad}$ such as
\begin{equation}
	\frac{ds_{rad}}{dt} = -3 H s_{rad} + \Sigma_{rad}\,,
\end{equation}
which can modify the relation between the temperature $T$ and the time $t$, resulting in a ``reheating'' of the primordial plasma and a local increase of the radiation entropy density. In absence of $\Sigma_{rad}$, this relation corresponds to $d(s_{rad} a^3)/dt=0$.

We use the following parametrisation:
\begin{equation}
	\Sigma_{rad}(T) = \kappa_{\Sigma_{r}} \Sigma_{rad}^{\rm eff}(T_{\rm BBN})\left( \frac{T}{T_{\rm BBN}} \right)^{n_{\Sigma_{r}}}\,,
\end{equation}
where $\kappa_{\Sigma_r}$ is the ratio of the radiation entropy injection to the radiation entropy density time-derivative $\Sigma_{rad}^{\rm eff}(T) \equiv \left| \dfrac{ds_{rad}}{dt}\right| = 3 H s_{rad}$ at $T_{\rm BBN} = 1$ MeV and $n_{\Sigma_r}$ is the decrease exponent of this production.

\subsubsection{Decaying scalar field}
\label{app:scalar}

We consider a scenario with a decaying pressureless primordial scalar field \cite{Gelmini:2006pw,Gelmini:2006pq}. The scalar field density $\rho_\phi$ is driven by the Boltzmann equation
\begin{equation}
 \frac{d\rho_\phi}{dt} = - 3 H \rho_\phi  - \Gamma_\phi \rho_\phi\,,
\end{equation}
where $\Gamma_\phi$ is the decay width of the scalar field. The scalar field decay results in radiation entropy injection such as:
\begin{equation}
\frac{ds_{rad}}{dt} = -3 H s_{rad} + (1-b) \frac{\Gamma_\phi \rho_\phi}{T}\,,
\end{equation}
where $b$ is the branching fraction of the decaying scalar field into supersymmetric particles. The non-thermal injection of supersymmetric particles affects the Boltzmann equation such as:
\begin{equation}
\frac{dn}{dt} = - 3 H n - \langle \sigma v \rangle (n^2 - n^2_{eq}) + \frac{b}{m_\phi} \Gamma_\phi \rho_\phi \,,
\end{equation}
with $m_\phi$ being the scalar field mass. One can define $\eta_\phi \equiv b/m_\phi$.

In general one can expect the relic density to be mainly generated thermally, so that $b/m_\phi$ remains small, and $b \ll 1$. The decay width can be related to the reheating temperature $T_{RH}$ through
\begin{equation}
 \Gamma_\phi = \sqrt{\frac{4\pi^3 g_{\rm{eff}}(T_{RH})}{45}} \, \frac{T^2_{RH}}{M_P}\,,
\end{equation}
where $g_{\rm{eff}}$ is the effective relativistic energy degrees of freedom, which can be obtained from the tables contained in \texttt{sgStar\_heff}.

This scenario requires three input parameters, the first one being $\tilde{\rho}_\phi$ the scalar field energy density divided by the photon energy density at the initial temperature (which can be either the freeze-out temperature or a user-defined temperature), the second one the reheating temperature $T_{RH}$ and the third one the value of $\eta_\phi$, which is expected to be small. $T_{init}$ should be sufficiently larger than the reheating temperature and freeze-out temperature in order to integrate properly the evolutions of the scalar field and of the supersymmetric particles.


\append{Indirect detection}
\label{app:indirect}
We turn to the calculation of the constraints from AMS-02 antiproton and Fermi-LAT gamma-ray data. These two types of constraint are derived from different kinds of analyses. However, they both require the calculation of the antiproton (or gamma-ray) flux produced from one dark matter annihilation. These two fluxes at production are computed following the procedure described below.
\subsection{Fluxes at production}
\label{subsec:fluxatprod}
Dark matter annihilates into pairs of Standard Model particles which subsequently hadronise into high-energetic cosmic rays. The flux at production of antiprotons or gamma rays can be expressed as the sum over dark matter annihilation channels $\chi \chi \to p_3 p_4$ of the antiproton ($\gamma$-ray) flux resulting from the hadronisation of particles $p_3$ and $p_4$ with an energy of $E_{p_3}$ and $E_{p_4}$ respectively, weighted by the channel branching ratio:
\begin{equation}
\frac{dN_{prod}}{dK}(K)= \sum_{\chi \chi \to p_3 p_4} BR(\chi \chi \to p_3 p_4)\left( \frac{dN_{p_3}}{dK}(E_{p_3}, K)  +\frac{dN_{p_4}}{dK}(E_{p_4}, K) \right)\ ,
\end{equation}
where $K$ is the kinetic energy of antiprotons ($\gamma$ rays).\\ 

Keeping in mind that the center of mass energy of dark matter annihilation processes is $\sqrt{s}= 2 m_{\chi}$, the energies $E_{p_3}$ and $E_{p_4}$ can be calculated from the energy and momentum conservation:
 
\begin{align}
\left \{
\begin{aligned}
E_{p_3}+E_{p_4}&=\sqrt{s}\,,\\
\overrightarrow{p_3}+ \overrightarrow{p_4}&= \overrightarrow{0}\,,
\end{aligned}
\right.
\end{align}
which gives
\begin{align}
\left \{
\begin{aligned}
E_{p_3}&=\frac{\sqrt{s}}{2} + \frac{M_3^2- M_4^2}{2 \sqrt{s}}\,,\\
E_{p_4}&=\frac{\sqrt{s}}{2} + \frac{M_4^2- M_3^2}{2 \sqrt{s}}\,,
\end{aligned}
\right .
\label{eq:e3e4}
\end{align}
where $M_3$ and $M_4$ are the outgoing particle masses.
In the case $M_3=M_4$, we retrieve the simple relation $E_{p_3}=E_{p_4}= \dfrac{\sqrt{s}}{2}= m_{\chi}$.\\

We use the tabulated spectra at production $\frac{dN^{PPPC4DMID}_{\chi \chi \to p \overline{p}}}{dK} (m_{\chi}=E_{p}, K)$ from the \texttt{PPPC4DMID} \cite{Cirelli:2010xx,Ciafaloni:2010ti}, which gives the flux of antiprotons ($\gamma$ rays) produced by one annihilation process of dark matter particles of a given mass $m_{\chi}$ annihilating via one of the following channels:
\begin{equation*}
\chi \chi \to e^{+} e^{-}, \ \mu^{+} \mu^{-}, \tau^{+} \tau^{-}, q \overline{q}, \ c \overline{c}, \ b \overline{b}, , \ t \overline{t}, \ W^{+} W^{-}, \ gg, \ \gamma \gamma , \ \nu_e \bar\nu_e, \ \nu_\mu \bar\nu_\mu, \ \nu_\tau \bar\nu_\tau \ ,
\end{equation*}
where $q$ stands for light quarks $u$, $d$, $s$. \\
These channels do not cover all the possible annihilation processes in the MSSM, so they cannot be used directly. However, it is possible to deduce the fluxes produced by the hadronisation of a SM particle $p$ with energy $E_p$ from these tabulated spectra: 
\begin{align}
\frac{dN_p}{dK}(E_p,k)=\frac{1}{2} \frac{dN^{PPPC4DMID}_{\chi \chi \to p \overline{p}}}{dK} (m_{\chi}=E_{p}, K)\,.
\end{align}

The spectra for the annihilation into light SM Higgs bosons of mass 125 GeV are also provided, however Higgs branching ratios may suffer from significant corrections in the MSSM. The hadronisation spectra of the lightest Higgs boson is therefore re-calculated using the branching ratios computed with \texttt{HDECAY} \cite{Djouadi:1997yw} or \texttt{FeynHiggs} \cite{Heinemeyer:1998yj}

\begin{equation}
\frac{dN_h}{dK}(E_h, K)= \sum_{h\to p_3 p_4} BR(h \to p_3\, p_4)\left( \frac{dN_{p_3}}{dK}(E_{p_3}, K)  +\frac{dN_{p_4}}{dK}(E_{p_4}, K) \right) \, ,
\end{equation}
where $E_{p_3}$ and $E_{p_4}$ are calculated from Eq. (\ref{eq:e3e4}) with $\sqrt{s}=E_h$. The hadronisation spectra of heavier Higgs bosons are then calculated in a similar way.\\
 
Finally, for the calculation of the neutralino annihilation branching ratios and cross sections in the MSSM and NMSSM, we use the same routines as for the relic density calculation to compute the neutralino annihilation amplitudes  $\chi \chi \to p_3 p_4$. \\
For a Majorana dark matter particle, at small velocity limit, the annihilation cross sections can easily be calculated from these amplitudes, following: 
\begin{equation}
\left< \sigma v\right >_{\chi \chi \to p_{3} p_{4}}= \frac{|\mathcal{A}|_{\chi \chi \to p_{3} p_{4}}^{2}}{128 \pi m_{\chi}^{2}} \left[1- \frac{M_{3}^{2}+M_{4}^{2}}{2m_{\chi}^{2}} + \frac{(M_{3}^{2}-M_{4}^{2})^{2}}{16 m_{\chi}^{4}} \right]^{\frac{1}{2}} \, .
\end{equation}

\subsection{Constraints from Fermi-LAT dwarf spheroidal  galaxies}
\label{sec:fermi}
The analysis of Fermi-LAT dwarf spheroidal galaxies (dSphs) is based on \cite{Fermi-LAT:2016uux}. Fermi-LAT collaboration performs a binned Poisson maximum-likelihood analysis in order to deduce the dark matter constraints. The energy range is separated into 24 bins, logarithmically spaced from 500 MeV to 500 GeV. Tabulated log-likelihoods are provided by the collaboration for each dSph and energy bin \cite{GLASTurl}. These tables allow us to estimate the log-likelihood $\mathcal{L}^i_j$ for a dSph $i$ and energy bin $j$ as a function of the gamma-ray flux produced by dark matter annihilation. 

The flux produced by the dark matter halo of a dSph $i$ in the energy bin $\left[ E^j_{\rm min},E^j_{\rm max} \right]$ is calculated as
\begin{align}
\begin{aligned}
\Phi^i_j=& \frac{1}{4 \pi} \frac{\sigmav}{2 m_{DM}^{2}}  \times  J^i \times\int_{E_{min}^j}^{E_{max}^j} \left(\frac{dN_{\rm prod}}{d E_{\gamma}}\right)_{\text{channel}}d E_{\gamma} \,,
 \end{aligned}
\end{align}
where $\frac{dN_{\rm prod}}{d E_{\gamma }}$ is the gamma-ray flux at production calculated as described in section \ref{subsec:fluxatprod} and $J^i$ is the J-factor of the dSph defined as  $J= \int_{\Delta \Omega} \int_{l.o.s} \rho_{DM}^{2}(r(l)) dl d \Omega$ with
$\Delta \Omega$ the solid angle under which the dSph is seen. The J-factor of each dSph is either deduced from their observed kinematics or, when no measurements are available, from an empirical law which states that the J-factor scales as the inverse square of the distance of the dSph.\\
In order to calculate the log-likelihood for a given dSph, we sum the log-likelihoods of every energy bins. Then, we add a corrective term to take into account the uncertainties on the J-factor:
\begin{equation}\label{eqn:jfactor_likelihood}
  \mathcal{L}^i(J_i) = \sum_{j} \mathcal{L}^i_{j}(J_i)-\frac{\left(log_{10}(J_i)-log_{10}(J_{obs,\, i})\right)^2}{2\sigma_i^2}\,,
\end{equation}
where $J_i$ is the true value of the J-factor, considered as a nuisance parameter, and $J_{obs,\,i}$ is the measured J-factor with error $\sigma_i$. For each dSph, a maximum log-likelihood is then calculated according to the nuisance parameters. Finally, we sum the maximum log-likelihood of every dSphs:
\begin{equation}
\mathcal{L}= \sum_{i}  \underset{J_i}{\rm max} \ \mathcal{L}^{i}(J_i) \,.
\end{equation}
The statistical test that we use to derive constraints on dark matter is then calculated by subtracting from the maximum log-likelihood of every dSphs the log-likelihood in the case where no dark matter is assumed:
\begin{equation}
{\rm TS}= 2 \left( \mathcal{L}_{\rm DM} -\mathcal{L}_{\rm no\, DM} \right) \ .
\end{equation}
This quantity follows a normal distribution and we will exclude points with a statistical test ${\rm TS}< \chi^2_0$, with $\chi^2_0$ a critical value depending on the desired confidence level and on the number of degrees of freedom. We choose the number of degrees of freedom (d.o.f.) as the number of annihilation channels which contribute at least to 1\% of the total annihilation cross section.


As already mentioned, the J-factor of some dSphs are calculated using an empirical relation. In order to assess the uncertainties on the log J-factor, we use 3 different values of $\sigma_i=$ 0.4, 0.6 or 0.8 dex for the \enquote{stringent}, \enquote{standard} and \enquote{conservative} options. \\
In addition, three different samples of dSphs are defined in Fermi-LAT analysis: a "conservative", a "nominal" and an "inclusive" sample, depending on the ambiguity of the kinematics of the galaxies. The "conservative" sample does not necessarily leads to a weaker limit compared to the "inclusive" one, as some dSphs in the "nominal" and "inclusive" samples show small but not significant enough excesses. Therefore, the delta log-likelihood is calculated for each sample. Our \enquote{conservative} option uses the largest delta log-likelihood, the \enquote{standard} option, the second largest and the \enquote{stringent} option, the smallest.

\subsection{Constraints from AMS-02 antiprotons} 
Contrary to gamma rays, antiprotons are diffused by turbulent magnetic fields in the galaxy. Therefore, it is necessary to describe the propagation of cosmic rays in the galactic medium in order to deduce the antiproton flux reaching Earth. To this end, we integrated in \texttt{SuperIso Relic} a code developed by Pierre Salati and Mathieu Boudaud, which is detailed  in \cite{Boudaud:2014qra}. The propagation model used in this code is briefly described below.

\subsubsection{Propagation Model}
We use a two-zone diffusion model where the galactic medium is a thin disk of $h=100pc$ height and $R=20$ kpc radius and cosmic rays are diffused a cylinder of half-height $L=$ and radius $R$. 

The antiproton spectrum respects the differential equation of propagation:
\begin{equation}
\frac{\partial f}{\partial t} - \mathcal{K}(K) \! \cdot \! \nabla^2f +
\frac{\partial}{\partial z}\left\{ {\rm sign}(z) \, f \, V_{\rm conv} \right\} +
\frac{\partial}{\partial E}\left\{ b(K , \vec{x}) f - \mathcal{K}_{\rm EE}(K) \frac{\partial f}{\partial E} \right\} = Q \, ,
\label{eq:transport}
\end{equation}
where $f=\frac{dN}{dK}(r,\, z,\, E)$ is the antiproton spectrum at radius $r$ and height $z$. We assume cylindrical symmetry, which allows us to decompose $f$ into Bessel transforms.
This method allows us to solve semi-analytically the equation of transport, which reduces significantly the computation time compared to a full numerical approach.\\
The first term in the equation of transport (\ref{eq:transport}) is set to zero since we only focus on steady-state solutions. 

\paragraph{Space diffusion:} 
The second term in Eq. (\ref{eq:transport}) describes the antiproton space diffusion with a coefficient
\begin{equation}
\mathcal{K}(K)=\mathcal{K}_{0} \beta p^{\delta} \, ,
\end{equation}
where $\beta=\frac{v}{c}$ is the antiproton beta factor and $p$ its momentum. $\mathcal{K}_{0}$ and $\delta$ are the free parameters of the model which set the normalisation and momentum dependence of the diffusion coefficient.\\
The third term in Eq. (\ref{eq:transport}) stands for convection processes, with a characteristic velocity $V_{conv}$ in the outside of the galactic disk. These processes tend to push the antiprotons vertically outside the disk.\\

\paragraph{Energy losses:}
The term in $b(K , \vec{x})$ accounts for energy losses. Antiprotons undergo energy losses according to three main processes: through the ionisation of the interstellar neutral medium, through the scattering off thermal electrons in interstellar ionised matter, and through convective processes. In addition, the inelastic but non-annihilating interactions of antiprotons with the interstellar medium (tertiary component) are treated as in~\cite{Boudaud:2014qra}.\\

\paragraph{Diffusive re-acceleration:}
Finally, the last term on the left-hand side of Eq. (\ref{eq:transport}) describes diffusive re-acceleration. The knots of the turbulent magnetic field can, in fact, drift the antiprotons with a characteristic velocity $v_a$, which results in a second order Fermi acceleration of antiprotons.\\

This model presents in total five free parameters which can be defined at the convenience of the user. However, the three benchmark propagation models \MIN{}, \MED{} and \MAX{}, which give a minimum, median and maximum antiproton flux at Earth, are directly provided (see table \ref{tab:propa}).

The term on the right-hand side of equation \ref{eq:transport} stands for the sources of antiprotons and will be detailed below.

\begin{table}[t!]
\begin{center}
\begin{tabular}{|c||c|c|c|c|c|}
\hline
Model		& $\delta$			& $\mathcal{K}_0$ [kpc$^2$/Myr]		& $L$ [kpc]		& $V_{conv}$ [km/s]	& $V_a$ [km/s]\\
\hline 
\hline
\MIN  		& 0.85			& 0.0016					& 1				& 13.5			&  22.4 \\
\MED  		& 0.70			& 0.0112					& 4 	 		& 12	 		& 52.9 \\
\MAX  		& 0.46			& 0.0765					& 15 			& 5	 			& 117.6\\
\hline
\end{tabular}
\caption[Benchmark \MIN, \MED, and \MAX\ sets of propagation parameters]{Benchmark \MIN, \MED, and \MAX\ sets of propagation parameters \cite{2004PhRvD..69f3501D}.}
\label{tab:propa}
\end{center}
\end{table}
\vspace*{0.3cm}

\subsubsection{Source terms}

\paragraph{Secondary antiprotons:} The astrophysical antiproton background, the so-called \emph{secondary antiprotons}, is mostly created through the interaction of proton and helium cosmic rays produced by supernovae with hydrogen and helium atoms in the interstellar medium. Such antiprotons account for one part of the source term in the equation of transport (\ref{eq:transport}):

\begin{equation}
Q_{\pbar}^{\rm II}(\mathbf{x} , K_{\pbar})  =  4 \pi  \, (1 + N_{\rm IS}(K_{\pbar})) {\sum_{i = {\rm p, He}}}  \; {\sum_{j = {\rm H, He}}}  {\displaystyle \int_{K^{0}_{i}}^{+ \infty}} dT_{i} \,
{\displaystyle \frac{d\sigma_{ij \to {\pbar} X}}{dK_{\pbar}}}(K_{i} \! \to \! K_{\pbar}) \, n_j (\mathbf{x}) \, \Phi_{i}(\mathbf{x} , K_{i}) \;,
\label{eq:Q_secondary}
\end{equation}
where $K_i$ is the kinetic energy of the nucleon $i$. The differential cross section $d\sigma_{ij \to {\pbar} X}/{dK_{\pbar}}$ is computed from the proton-proton differential cross section taken from ReF.~\cite{diMauro:2014zea} and the threshold $K_p^0$ of this reaction is taken to be $7m_p$.

The factor $N_{\rm IS}$ accounts for the fact that antineutrons are also produced, along with antiprotons, in the interaction of cosmic rays with the interstellar medium. Antineutrons subsequently decay into antiprotons and contribute to the source term. Most of the antiprotons are, however, produced in proton-proton reactions.

The flux of proton and helium cosmic rays, at position $\mathbf{x}$, $\Phi_{i}(\mathbf{x})$ are deduced from their fluxes measured at the position of the Earth through a retropropagation technique.

\paragraph{Primary antiprotons:}
The production rate $Q_{\pbar}^{\rm I}$  of primary antiprotons produced by the annihilation of two dark matter particles is given by:

\begin{equation}
Q_{\pbar}^{\rm I}(\mathbf{x} , K_{\pbar}) = \eta \left( \frac{\rho(\textbf{x})}{m_{\rm DM}} \right)^2  \sigmav \frac{dN_{prod}}{dK_{\pbar}}\;,
\label{eq:Q_primary}
\end{equation}
where $\sigmav$ is the thermal average annihilating cross section, and $\eta$ is equal to $1/2$ ($1/4$) for a Majorana (Dirac) type particle. $\dfrac{dN_{prod}}{dK_{\pbar}}$ is the flux at production of antiprotons which is calculated as described in section \ref{subsec:fluxatprod}.

\subsubsection{Calculation of the constraints}
We calculate the total antiproton spectrum at the Earth position as the sum of primary and secondary antiproton contributions:
\begin{equation*}
\Phi^\oplus_{\rm tot}(K, \phi_F, A)= \Phi^\oplus_{\rm I}(K, \phi_F) + \Phi^\oplus_{\rm II}(K, \phi_f, A) \ .
\end{equation*}
The parameters $A$ and $\phi_F$ are  nuisance parameters over which we marginalise. They are related to the uncertainties on the antineutron production cross section and to the solar modulation, respectively.

\paragraph{Uncertainties on the antineutron production cross section:}
The parameter $N_{\rm IS}$ which accounts for antineutron production in Eq. (\ref{eq:Q_secondary}) is energy-dependent and suffers from large uncertainties \cite{diMauro:2014zea}. 
For this reason, we calculate the secondary antiproton spectra for the lower and upper bound of $N_{\rm IS}$. The real secondary antiproton spectrum takes values between these two bounds, according to a nuisance parameter $A\in [0,1]$
\begin{equation}
\Phi_{\rm II}(A)=(1-A) \times \Phi_{\rm II}^{lower} + A\times \Phi_{\rm II}^{upper} \,.
\end{equation}

\paragraph{Solar modulation:}

As antiprotons reach the Sun vicinity, they enter the sphere of influence of the Sun magnetic field and of its cosmic-ray wind. It has for effect to decrease the kinetic energy of antiprotons, especially for low-energetic particles ($ \lesssim 10$ GeV). It is therefore necessary to modify the spectrum at the Earth position calculated from the equation of transport. A simple way to do this is to use a force-field approximation parametrised by the Fisk potential $\phi_F$ \cite{Gleeson:1967juf,Gleeson:1968zza}:

\begin{equation}
\Phi^\oplus(K)=\Phi^{\rm 0}(K+ |e| \phi_F )\cdot \frac{ K(K+2m_{p})}{(K+m_{p}+ |e| \phi_F)^2-m_{p}^2}.
\label{eq:SMod}
\end{equation}

where $\Phi^{\oplus}$ is the antiproton spectrum at Earth and $\Phi^{\rm 0}$ the antiproton spectrum at the end of the propagation but before entering into the solar influence. For AMS-02 data, we take $\phi_F \in [0.1,1]$.\\

In order to quantify the deviation of the theoretical spectrum from the antiproton spectrum measured by AMS-02, we calculate a $\chi^2$ as: 

\begin{equation}
\chi^{2}(\Phi^\oplus_{\rm tot})=\sum_{i} \left(\frac{\Phi^\oplus_{\rm tot}(E_i)-\Phi^\oplus_{AMS-02}(E_i)}{\Delta\Phi^\oplus_{AMS-02}(E_i))}\right)^{2}\ ,
\end{equation}
where we sum over AMS-02 energy bins, with central values $E_i$. AMS-02 flux $\Phi^\oplus_{AMS-02}$ is given with an error $\Delta\Phi^\oplus_{AMS-02}$ \cite{Aguilar2015c_bis,Aguilar2015b_bis}.\\
We minimise the $\chi^2$ with respect to the nuisance parameters $A$ and $\phi_F$ and we compare it to the $\chi^2$ of the case without dark matter: 
\begin{equation}
\Delta\chi^{2}= \underset{A,\phi_F}{min} \left\{ \chi^{2}(\Phi^\oplus_{\rm tot})\right\} - \underset{A,\phi_F}{min}\left \{ \chi^{2}(\Phi^\oplus_{\rm no DM})\right \} \,.
\end{equation}
The obtained $\Delta\chi^{2}$ follows a normal distribution. For instance, if only one annihilation channel is dominant (1 d.o.f.), a point will be excluded at 2 $\sigma$ if $\Delta\chi^{2}>4$.\\

The computation of antiproton primary and secondary spectra can be time-consuming when performing large scans. This is the reason why we also provide tabulated spectra for the benchmark sets of propagation parameters \MIN{}, \MED{}, \MAX{}, and the three DM halo density profiles Burkert, Einasto and Navarro-Frenk-White (NFW), defined in table \ref{tab:haloprofiles}. Following our analysis of AMS-02 antiproton constraints in the MSSM \cite{Arbey:2017eos}, we define the limit derived by the Burkert profile and MED propagation model as the \enquote{conservative} constraint, whereas the Einasto \MED{} and the Einasto \MAX{} models provide \enquote{standard} and \enquote{stringent} constraints, respectively.\\

\begin{table}[t!]
\centering
\begin{tabular}{|l|c|c|c|c|}
\hline
Halo profile &$r_s$ [kpc] &$\rho_{s}$[GeV/cm$^3$]& $R_{\odot}$ [kpc]& $\rho_{\odot}$[GeV/cm$^3$]\\
\hline
NFW&$19.6$&$0.32$&$8.21$&$0.383$\\
Einasto ($\alpha=0.22$) &$16.07$ &$0.11$&$8.25$&$0.386$\\
Burkert&$9.26$&$1.57$&$7.94$&$0.487$\\
\hline
\end{tabular}
\caption[Dark matter mass model parameters for NFW, Einasto and Burkert profiles]{Dark matter mass model parameters for NFW~\cite{2011MNRAS.414.2446M}, for Einasto~\cite{Catena:2009mf} and for Burkert~\cite{Nesti:2013uwa} profiles.}
\label{tab:haloprofiles}
\end{table}

However, the users are completely free to define their own propagation parameters and DM density profile, as long as it respects axisymmetry. The constraints can, in this way, be calculated directly for the new sets of parameters. If the user wishes to perform large scans, we advise, nevertheless, to generate tabulated spectra of primary and secondary antiprotons using the routines \texttt{void write\_secondaries} and \texttt{void write\_primaries}.


\append{Direct detection}
\label{app:direct}
\section{Direct detection}
\subsection{Generalities}
The calculation of direct detection constraints requires the computation of the differential recoil rate per unit of target material mass $\dfrac{dR}{dE}$:
\begin{equation}\label{eqn:dRdEnr1}
  \frac{dR}{dE}
    = \frac{n_{\chi}}{M} \, \Big\langle v \frac{d\sigma}{d E} \Big\rangle
    = \frac{2\rho_{\chi}}{m_{\chi}}
      \int d^3v \, v f(\mathbf{v},t) \, \frac{d\sigma_{i}}{dq^2}(q^2,v) \, ,
\end{equation}
where $n_{\chi} = \rho_{\chi}/m_{\chi}$ is the number density of \wimps{}, with $\rho_{\chi}$ the local DM mass density. $f(\mathbf{v},t)$ is the \wimp{} velocity distribution and $\dfrac{d\sigma_{i}}{dq^2}(q^2,v)$ is the differential
\wimp{}/nucleus cross section, with $q^2 = 2 M E$ the momentum exchanged in the scattering.
Using the form of the differential cross section for the most
commonly assumed couplings, Eq. (\ref{eqn:dRdEnr1}) can be simplified as
\begin{equation}\label{eqn:dRdEnr2}
  \frac{dR}{dE}
    = \underbrace{\frac{1}{2 m_{\chi} \mu^2}\; \sigma(q)}_{\text{Particle physics}}  \; \underbrace{\vphantom{\frac{1}{m_{\chi}}}\rho_{\chi} \eta(v_{min}(E),t)}_{Astrophysics} \, ,
\end{equation}
 where $\sigma(q)$ is an effective scattering cross-section and
\begin{equation} \label{eqn:eta}  
  \eta(v_{min},t) \equiv \int_{v > v_{min}} d^3v \, \frac{f(\mathbf{v},t)}{v}
\end{equation}
is the mean inverse speed. $v_{min}$ can be written as
\begin{equation} \label{eqn:vmin} 
  v_{min} =\sqrt{\frac{M  E}{2\mu^2}}\,,
\end{equation}
where $\mu $ is the reduced mass $\mu = m_{\chi} M/ (m_{\chi} + M)$.
In the general case where the target material is composed of more than one isotope, the total differential recoil rate is the sum of each isotope contribution weighted by the isotope mass fraction:
\begin{equation}
  \frac{dR}{dE}= \sum_{i } \xi_{i} \frac{dR_{i}}{dE} \ .
  \end{equation}
One can note that the first part in the right side of Eq. (\ref{eqn:dRdEnr2}), in which the effective cross section appears, depends only on the particle model which is used. The main source of uncertainties in this term comes from nuclear form factors. The second part involves only astrophysical observables, namely dark matter local density and velocity profile. The calculation of each term is, therefore, to be done separately.\\

The calculation of $\eta(v_{min},t)$ is performed with a Simpson method. We use the standard halo model which describes the dark matter halo as a non-rotating isothermal sphere \cite{Freese:1987wu, Drukier:1986tm}. It involves an isotropic Maxwellian distribution of the \wimp{} velocity $f(\mathbf{v})$, with the galactic disk rotation velocity $v_{rot}$ being the most probable speed. We set the default $v_{rot}$ central value and uncertainties to $v_{rot}=220 \pm 20$ km/s \cite{Reid:2009nj,McMillan:2009yr,Bovy:2009dr}.
This velocity distribution is truncated at the escape velocity $v_{esc}$ at which a \wimp{} can escape the galaxy potential well, $v_{esc}=544 \pm 50$ km/s \cite{Smith:2006ym}. DM direct detection constraints are, however, mostly influenced by the large uncertainties on the local DM density $\rho_{0}=0.4 \pm 2$ GeV/cm$^3$ \cite{Read:2014qva}.
A \enquote{conservative}, \enquote{standard}, or \enquote{stringent} constraint can be calculated with respect to the uncertainties of these three parameters. We refer to ref. (\cite{Arbey:2017eos}) for the study of the impact of these uncertainties on direct detection constraints.

\subsection{Scattering cross sections}
The effective \wimp{}/nucleus scattering cross section is commonly decomposed into a spin-dependent and spin-independent component $\sigma_{i}(q)=\sigma^{SI}_{i}(q)+\sigma^{SD}_{i}(q)$. Each component is the subject of a special treatment, but both need the calculation of effective neutralino-quark couplings which will be detailed, in the MSSM, in section  \ref{subsubsec:couplings}.\\

\subsubsection{SI and SD cross sections}
The \textbf{spin-independent} effective cross section can be written for a target nucleus composed of $Z$ protons and $(A-Z)$ neutrons as:
\begin{equation}
\sigma^{SI}_{i}(q)=\frac{4 \mu^{2}}{\pi} F_{SI}^{2}(q) \left[Z \times A_{p}^{SI} +(A-Z) \times A_{n}^{SI}  \right]^{2}\,,
\end{equation}
where $F_{SI}$ is a nuclear form factor which probes the nucleon content of the nucleus, and $A_{p}^{SI}$ and $A_{n}^{SI}$ are the proton/\wimp{} and neutron/\wimp{} effective scattering amplitudes, respectively. These amplitudes can be calculated from the \wimp{}/quark effective couplings $\lambda^{SI}_{q}$ weighted by the quark form factors.
\begin{equation}
A_{p,\, n}^{SI}= \sum_{q=u,d,s} f_{q}^{p,\, n} \lambda^{SI}_{q} \ .
\end{equation} 
The quark form factors $f_{q}^{p,\, n}$ probe the mass content of the quarks in the nucleons:
\begin{equation}
f_{q}^{N}= \frac{m_q}{M_N}\big \langle N| \overline{\Psi}_q \Psi_q    |N \big \rangle \ .
\end{equation} 
As for the \textbf{spin-dependent} counterpart, the effective cross section can be written as:
\begin{equation}\label{eqn:dsigmadqS}
\sigma_{i}^{SD}(q)=\frac{16 \mu^2}{2J+1} S(q) \ ,
\end{equation}
where $J$ is the total spin of the nucleus and 
\begin{equation}
  S(q)=a_0^2\,S_{00}(q)+a_1^2\,S_{11}(q)+a_0\;\!a_1\;\!S_{01}(q)\; .
\end{equation}
$S_{00}$, $S_{01}$ and $S_{11}$ are structure factors which depend on the nature of the target isotope and $a_0$ and $a_1$ are defined as
\begin{equation}
  a_0=A_{p}^{SD}+A_{n}^{SD} \,,\qquad a_1= A_{p}^{SD}-A_{n}^{SD}\;,
\end{equation}
where $A_{p,\,n}^{SD}$ are the \wimp{}/proton or neutron spin-dependent scattering amplitudes. In the same way as for the spin-independent amplitude, $A_{p,\,n}^{SD}$ are calculated as the sum of the \wimp{}/quark SD effective couplings $\lambda^{SD}_{q}$, weighted by quark form factors probing the spin content of the quarks in the nucleons $\Delta q^{p,\,n}$.
\begin{equation}
A_{p,\,n}^{SD}=\sum_{q=u,d,s} \Delta q^{p,\,n} \lambda^{SD}_{q} \ .
\end{equation}
$\Delta q^{p,\,n}$ is defined as:
\begin{equation}
 \Delta q^{N}= \frac{1}{2 s_{\mu}} \big \langle N| \overline{\Psi}_q  \gamma_{\mu} \gamma^{5} \Psi_q    |N \big \rangle \ ,
\end{equation}
where $s_{\mu}$ is the nucleon spin.

\subsubsection{Neutralino/quark effective couplings}
\label{subsubsec:couplings}
An important step in the calculation of the scattering cross sections consists in computing the effective couplings $\lambda_{q}^{SI}$ and $\lambda_{q}^{SD}$.  To this end, we use the explicit analytical formulae from Ref.~\cite{Drees:1993bu}
for the calculation of the SI and SD effective couplings in the MSSM and in the NMSSM.
The implementation of the neutralino-nucleon scattering amplitudes in the NMSSM requires only a small modification of the Higgs couplings, as the additional CP-even Higgs also participates to the scattering process. These couplings were calculated using FormCalc \cite{Hahn:2016ebn}.
We treat the twist-2 operator and the box diagrams as described in Ref. \cite{Drees:1993bu}.\\
Additional higher-order corrections were implemented for the spin-independent amplitude. The neutralino can indeed interact with the gluon content of the nucleons via diagrams including heavy quarks or squark loops.
At first approximation, these diagrams can be treated as effective interactions between the neutralino and the heavy quarks and squarks, corrected by an appropriate form factor $f_Q$ related to the gluon content of the nucleon. For heavy quarks, this form factor depends simply on the sum of the light quark form factors:

\begin{equation}
f_{Q}^{p,\,n}= \frac{2}{27}\left( 1- \sum_{q=u,d,s} f_{q}^{p,\,n}\right)\,.
\end{equation}
This form factor is however enhanced in the case of a Higgs boson exchange due to QCD effects and we apply the corrections given in Ref.~\cite{Djouadi:2000ck}.

Finally, the bottom quark mass may receive significant corrections from gluino-squark loop in SUSY scenarios. While down-type quarks couple normally only to the Higgs doublet $H_d$, this correction allows also a small coupling between the $b$-quark and $H_u$. This SUSY-QCD correction modifies the coupling of the $b$-quark to CP-even Higgs as described in Ref.~\cite{Djouadi:2000ck}.

\subsection{Uncertainties on the  nucleon and nuclear form factors}

\subsubsection{Nucleon form factors}

\paragraph{Spin-independent interaction}

Starting with the spin-independent nucleon form factors,  $f^{p,\, n}_{u}$, $f^{p,\, n}_{d}$ and  $f^{p,\, n}_{s}$ can be calculated from three parameters, namely the light and strange quark contents of the nucleon, defined as
\begin{align}
\Sigma_{\pi N} &\equiv \frac{m_u+m_d}{2} \langle N | \overline{u}u + \overline{d}d | N \rangle \ ,\\
\sigma_s &\equiv m_s \langle N | \overline{s}s | N \rangle \, ,
\end{align}
and the parameter $z$ defined as:
\begin{equation}
z=  \frac{\big \langle N|\overline{u}u |N \big \rangle -\big \langle N|\overline{s}s |N \big \rangle}{\big \langle N|\overline{d}d |N \big \rangle -\big \langle N|\overline{s}s |N \big \rangle} \,.
\end{equation}
The light quark content of the nucleons is known from lattice QCD results and from the analysis of pion-nucleon scattering. As for the strange content of the nucleons, it is deduced from lattice QCD calculations. We use the range of values of $\Sigma_{\pi N}$, $\sigma_s$ and $z$ from Ref.~\cite{Ellis:2018dmb}:
\begin{align}
\sigma_l&=46 \pm 11\ {\rm MeV} \ , \\
\sigma_s&=35 \pm 16\ {\rm MeV} \ ,\\
z&= 1.5 \pm 0.5 \ .
\end{align}

\paragraph{Spin-dependent interactions}

The nucleon SD form factors $\Delta^{(N)}_q$, can be computed from the combinations of two parameters:
\begin{eqnarray}
a_3 &=& \Delta^{(p)}_u - \Delta^{(p)}_d  \ , \\ 
a_8 &=& \Delta^{(p)}_u + \Delta^{(p)}_d - 2 \Delta^{(p)}_s   ,
\end{eqnarray}
and $\Delta^{(p)}_s$ itself.
$a_3$ is known from neutron $\beta$ decays measurements 
$a_3 = 1.2723 \pm 0.0023$ \cite{Agashe:2014kda} and $a_8 = 0.585 \pm 0.023$ from hyperon $\beta$ decay results \cite{Goto:1999by}. The uncertainties on these two parameters are rather small compared to the error on $\Delta^{(p)}_s = -0.09 \pm 0.03$, which is deduced from the measurement of the spin-dependent structure function of the deuteron from the COMPASS experiment \cite{Alexakhin:2006oza}.

\subsubsection{Nuclear structure factors}
\label{subsubsec:nuclear}

\paragraph{Spin-independent interactions}

The SI form factor is essentially a Fourier transform of the mass distribution of the nucleus. A good approximation is
the Helm form factor \cite{Helm:1956zz,Lewin:1995rx}:
\begin{equation} \label{eqn:SIFF}
  F(q) = 3 e^{-q^2 s^2/2} \; \frac{\sin(qr_n)- qr_n\cos(qr_n)}{(qr_n)^3} \, ,
\end{equation}
where $s\simeq 0.9$~fm and $r_n^2 = c^2 + \frac{7}{3} \pi^2 a^2 - 5
s^2$ is an effective nuclear radius with $a \simeq 0.52$~fm and $c
\simeq 1.23 A^{1/3} - 0.60$~fm. The uncertainties on the Helm form factors are expected to be rather small for small momentum exchanges. However, the users may define their own SI form factor if desired.

\paragraph{Spin-dependent interactions}

Large uncertainties on $S_{01}$ and $S_{11}$ structure factors exist from long-range two-body currents due to pion exchange \cite{Klos:2013rwa}. There is a correlation between $S_{01}$ and $S_{11}$ errors, so one should be aware that we may slightly overestimate the uncertainties on $S(q)$ by taking independently the lower and upper bounds of $S_{01}$ and $S_{11}$ to estimate the impact of the uncertainties. However, this method allows us to stay conservative.\\
The uncertainties are especially relevant for isospin violating models, since $S_{01}$ and $S_{11}$ are proportional to $A_p^{SD}-A_n^{SD}$. 

.

\subsection{Experimental limits}
We implemented the constraints from the latest results of XENON1T \cite{Aprile:2017iyp}, PANDAX-2 \cite{Cui:2017nnn} and PICO60 \cite{Amole:2017dex}. No events were observed during the time of exposure of these experiments, except for PANDAX-2, which observed one event during RUN 9.
For each constraint, we use a Poisson likelihood defined as:

\begin{equation}
\mathcal{L}\left( \mu | N_o \right)= \frac{(b+\mu)^{N_o} e^{-(b+\mu)}}{N_o !} \ ,
\end{equation}
where $\mu$ is the total expected number of events from \wimp{}-nucleus scattering, $N_o$ is the number of observed events and $b$ is the expected background. We marginalise over $b$ by choosing its value according to:
\begin{align}
\left \{
\begin{aligned}
b&=N_o -\mu \qquad &\text{if }N_o>\mu \,,\\
b&=0 & \text{otherwise.}
\end{aligned}
\right .
\end{align}

A point is excluded at 90\% C.L. if the difference of its log-likelihood with the background-only scenario respects:
\begin{equation}
\log\, \mathcal{L}(\mu=0) - \log\, \mathcal{L}(\mu)> \frac{2.71}{2} \ .
\end{equation}

We use the efficiencies provided in the GAMBIT package DDCalc \cite{Workgroup:2017lvb}. They were calculated using the \texttt{TPCMC} Monte Carlo code \cite{Savage:2015tpcmc} to model the detector response, which relies on \texttt{NEST} \cite{Szydagis:2011tk,Szydagis:2013sih,NEST:url} for modelling the microphysics of a recoiling xenon atom.

%
%
\newpage

\bibliographystyle{JHEP}
\bibliography{biblio}

\end{document}